\documentclass[fleqn,10pt]{wlscirep}
\title{
Universal adaptive optics for microscopy through embedded neural network control}

\author[1]{Qi Hu}
\author[2]{Martin Hailstone}
\author[1]{Jingyu Wang}
\author[1]{Matthew Wincott}
\author[2]{Danail Stoychev}
\author[3]{Huriye Atilgan}
\author[2]{Dalia Gala}
\author[2]{Tai Chaiamarit}
\author[2]{Richard M. Parton}
\author[1]{Jacopo Antonello}
\author[3]{Adam M. Packer}
\author[2]{Ilan Davis}
\author[1,*]{Martin J. Booth}
\affil[1]{Department of Engineering Science, University of Oxford}
\affil[2]{Department of Biochemistry, University of Oxford}
\affil[3]{Department of Physiology, Anatomy, and Genetics, University of Oxford}
\affil[*]{martin.booth@eng.ox.ac.uk}

\begin{abstract}
The resolution and contrast of microscope imaging is often affected by aberrations introduced by imperfect optical systems and inhomogeneous refractive structures in specimens. Adaptive optics (AO) compensates these aberrations and restores diffraction limited performance. A wide range of AO solutions have been introduced, often tailored to a specific microscope type or application. Until now, a universal AO solution -- one that can be readily transferred between microscope modalities -- has not been deployed. We propose versatile and fast aberration correction using a physics-based machine learning assisted wavefront-sensorless AO control (MLAO) method. Unlike previous ML methods, we used a bespoke neural network (NN) architecture, designed using physical understanding of image formation, that was embedded in the control loop of the microscope. The approach means that not only is the resulting NN orders of magnitude simpler than previous NN methods, but the concept is translatable across microscope modalities. We demonstrated the method on a two-photon, a three-photon and a widefield three-dimensional (3D) structured illumination microscope. Results showed that the method outperformed commonly-used modal-based sensorless AO methods. We also showed that our ML-based method was robust in a range of challenging imaging conditions, such as extended 3D sample structures, specimen motion, low signal to noise ratio and activity-induced fluorescence fluctuations. Moreover, as the bespoke architecture encapsulated physical understanding of the imaging process, the internal NN configuration was no-longer a ``black box'', but provided physical insights on internal workings, which could influence future designs.
\end{abstract}
\begin{document}

\flushbottom
\maketitle
%
%
\thispagestyle{empty}

\section*{Introduction}
The imaging quality of high-resolution optical microscopes is often detrimentally affected by aberrations which result in compromised scientific information in the images. These aberrations can arise from imperfections in the optical design of the microscope, but are most commonly due to inhomogeneous refractive index structures within the specimen. Adaptive optics (AO) has been built into many microscopes, restoring image quality through aberration correction by reconfigurable elements, such as deformable mirrors (DMs) or liquid crystal spatial light modulators (LC-SLMs).\cite{Booth2007,Booth2014,Booth2014a,booth2015,Ji2017,Hampson2021} Applications of AO-enabled microscopes have ranged from deep tissue imaging in multiphoton microscopy through to the ultra-high resolution required for optical nanoscopy. This range of applications has led to a wide variety of AO solutions that have invariably been tailored to a specific microscope modality or application. ikbn 

There are two main classes AO operation: in one case, a wavefront sensor measures aberrations; in the other case, aberrations are inferred from images -- so called ``wavefront sensorless AO", or ``sensorless AO" for short. For operations with a wavefront sensor, phase aberrations are measured directly by wavefront sensors such as a Shack-Hartmann sensor \cite{1904Hartmann,Shack1971} or an interferometer \cite{Schwertner2004,Booth:05
}. Such operations are direct and fast but also have intrinsic disadvantages such as requiring a complex optical design and suffering from non-common path errors. Furthermore, such wavefront sensors often have limitations and are less versatile. For example, an interferometer requires a coherent source and all such methods suffer from problems due to out-of-focus light. On the other hand, sensorless AO methods normally function with a simpler optical design and thus are more easily adaptable for a wide range of imaging applications. However, sensorless AO methods are based on iterative deductions of phase aberrations and thus tend to be more time consuming; this is coupled with repeated and prolonged sample exposures, which inevitably lead to photo-damage or motion related errors.

There have been many developments in AO technology, and in particular sensorless AO methods. Conventionally, sensorless AO operates based on the principle that the optimal image quality corresponds to the best aberration correction \cite{Hu2020,hu2021a}. A suitably defined metric, such as the total signal intensity \cite{Booth:02,Sherman02,Marsh:03,wright2005exploration,
Debarre2009,Tang2012,Facomprez2012,
Katz:14,
Sinefeld2015,Galwaduge:15,Streich2021} or a spatial frequency based sharpness metric \cite{Debarre:07,
Gould2012,Bourgenot:12,Burke:15,Patton:16}, is used to quantify the image quality. Phase is modulated by the AO while this quality metric reading is measured and optimised. There have been discussions on how the phase should be modulated \cite{
Wang,Milkie2011,
Hu2020} and how the optimisation algorithm should be designed \cite{Booth2002,Wang2009,Antonello2012,Facomprez2012}. However, as mentioned before, such ``conventional" sensorless AO methods depend on iterative optimisation of a scalar metric, where all image information is condensed into a single value, and the optimisation process is usually through mode by mode adjustment. Such methods were thus not the most efficient approach to solving this multi-dimensional optimisation problem and the effective range of correction was limited. While a higher dimensional metric was considered to extract more information from images \cite{Antonello:20}, the optimisation of such a vector metric was not straightforward.

While the utility of each of these conventional sensorless AO methods has been demonstrated separately, each method had been defined for a particular microscope type and application. Until now, no such AO solution has been introduced that can be universally transferred between microscope modalities and applications.  

We propose in this article a new approach to sensorless AO (named as MLAO) that addresses the limitations of previous methods and provides a route to a universal AO solution that is applicable to any form of microscopy. This solution is constructed around a physics-based machine learning (ML) framework that incorporates novel neural network (NN) architectures with carefully crafted training procedures, in addition to data pre-processing that is informed by knowledge of the image formation process of the microscope. The resulting NN is embedded into the control of the microscope, improving the efficiency and range of sensorless AO estimation beyond that possible with conventional methods. This approach delivers versatile aberration measurement and correction that can be adapted to the application, such as the correction of different types of aberration, over an increased range of aberration size, across different microscope modalities and specimens.

In recent years, machine learning (ML) has been trialed in AO for its great computational capability to extract and process information. However, many of these approaches required access to point spread functions (PSFs) or experimentally acquired bead images \cite{Jin:18,mockl2019,Vishniakou:20,Cumming:20,Khorin_2021,Zhang:22,Saha:20} 
; these requirements limited the translatability of these methods to a wider range of applications. Reinforcement learning was applied to correct for phase aberrations when imaging non point-like objects \cite{Durech:21}; however, the method still involved iterative corrections and was not advantageous in terms of its correction efficiency, accuracy and correction working range compared to conventional sensorless AO algorithms. Untrained neural networks (NN) were used to determine wavefront phase and were demonstrated on non point-like objects \cite{Wang2020,Bostan:20}; however, such methods were reported to normally require a few minutes of network convergence, which limits their potential in live imaging applications.

Our new approach differs considerably from previous ML assisted aberration estimation, as previous methods mostly employed standard deep NN architectures that used raw images as the input data. Our method builds upon physical knowledge of the imaging process and is designed around the abilities of the AO to introduce aberration biases, which improve the information content of the NN input data. This approach means that the resulting NN is orders of magnitude simpler, in terms of trainable parameters, than previous NN methods (See Table S1 in supplemental document). Furthermore, our method is readily translatable across microscope modalities. As NN training is carried out on a synthetic data set, adaptation for a different modality simply requires regeneration of the image data using a new imaging model. The NN architecture and training process are otherwise similar.

To illustrate the versatility of this concept, we have demonstrated the method on three different types of fluorescence microscopes with different forms of AO corrector: a two-photon (2-P) microscope using a SLM, a three-photon (3-P) intravital microscope using a DM, and a widefield three dimensional (3-D) structured illumination microscope (SIM) using a DM. In all cases, we showed that the new method outperformed commonly used conventional sensorless AO methods. The results further showed that the ML-based method was robust in a range of challenging imaging conditions, such as specimen motion, low signal to noise ratio (SNR), and fluorescence fluctuations. Moreover, as the bespoke architecture encapsulated into its design physical understanding of the imaging process, there was a link between the weights in the trained NN and physical properties of the imaging process.
This means that the internal NN configuration needs no-longer to be considered as a ``black box", but can be used to provide physical insights on internal workings and how information about aberrations is encoded into images. 

\begin{figure}
    \centering
    \includegraphics[width = 1.0\textwidth]{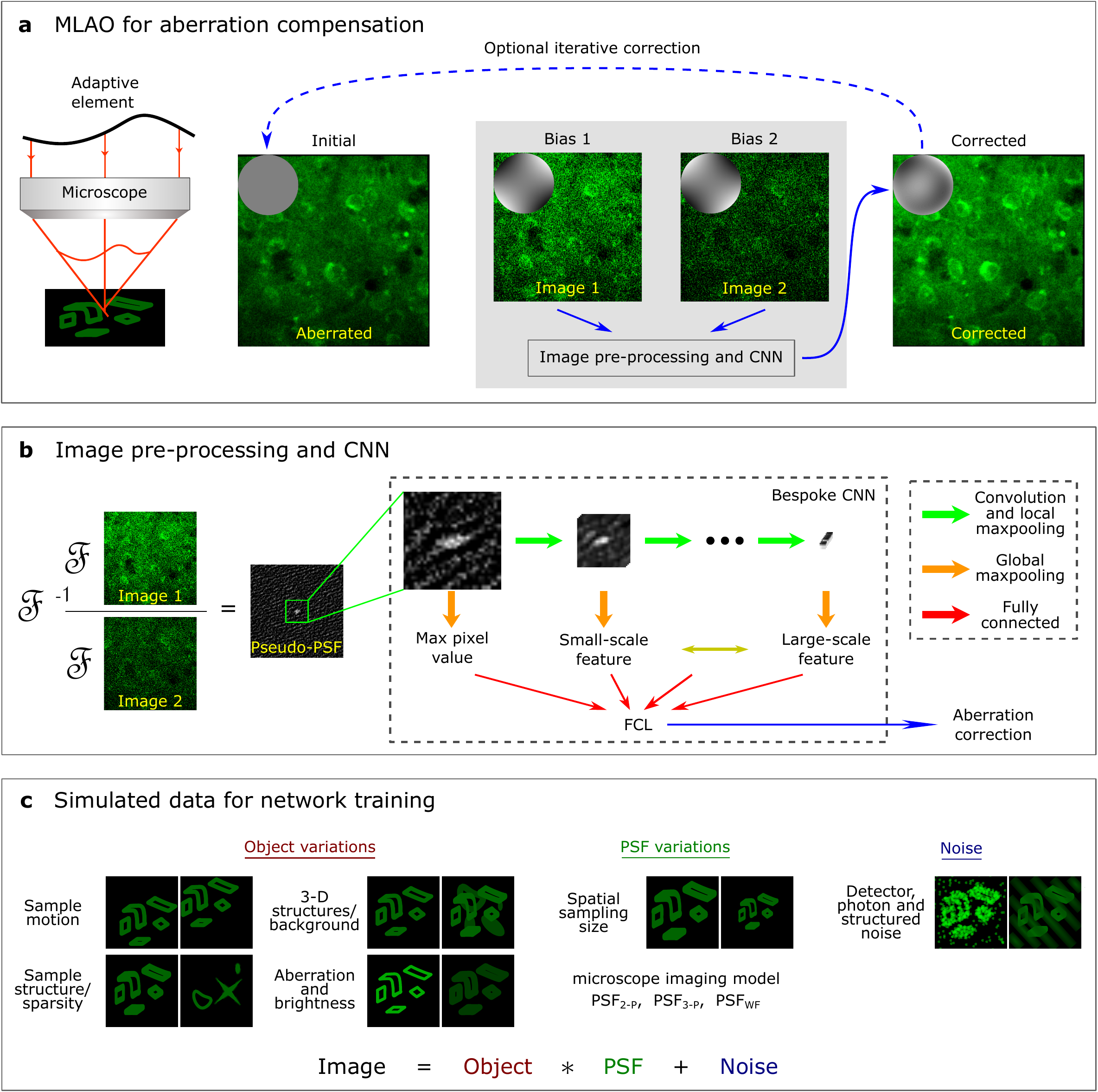}
    \caption{The MLAO concept. (a) Overview of the AO correction process. A minimum of two bias aberrations were introduced by the adaptive element; corresponding images of the same field were captured. The images were passed to the MLAO estimator, which determined the Zernike coefficients for correction. The correction speed was limited only by the speed of image acquisition, not by computation. Further correction could optionally be implemented through iteration.
    (b) Image pre-processing and NN architecture. Images were pre-processed to compute pseudo-PSFs, which were predominantly independent of specimen structure. $\mathcal{F}$ and $\mathcal{F}^{-1}$ represent the forward and inverse Fourier transform, respectively. A central cropped region of the pseudo-PSF images was used as the input to a CNN. The CNN  was designed and trained specifically for aberration determination. The output from the overall network was the correction coefficients for the Zernike modes. The NN architecture was such that the convolutional layer outputs could be correlated with spatial scales of the aberration effects on the pseudo-PSFs and hence the imaging process. Hence, the distribution of weights in the network had physical relevance.
    (c) Training data generation. A range of image variations were included in the synthetic data set for NN training to cope with variations in real experimental scenarios. The data were a combination of artificial and real microscope images, chosen to model a wide range of realistic specimen structures. Images were created through convolution of specimen structures with an appropriate PSF, generated for the specific microscope modality, incorporating aberrations.
    }
    \label{fig:MLAO_demonstration}
\end{figure}

\section*{Concept and implementation}\label{section:concept}

The overall MLAO concept is illustrated in Figure \ref{fig:MLAO_demonstration}. The  experimental application follows closely the concept of modal sensorless AO, whereby a sequence of images are taken, each with a different bias aberration applied using the adaptive element. The set of images are then used as the input to the ML-enabled estimator, which replaces the previous conventional method of optimisation of an image quality metric. The estimated correction aberration is then applied to the adaptive element. If necessary, the process can be iterated for refined correction. The significant advantage of the new method is the way in which the estimator can more efficiently use image information to determine the aberration correction.

The concept has been designed in order to achieve particular capabilities that extend beyond those of conventional sensorless AO. The new method should ideally achieve more efficient aberration estimation from fewer images, to reduce time and exposure of measurement.
It should operate over a larger range of aberration amplitudes, compared to previous methods. A particular estimator should be robust to variations between similar microscopes and the concept should be translatable across different microscope types and applications. From a practical perspective, it is also important that training can be performed on synthetic data, as it would be impractical to obtain the vast data set necessary for training from experimentally obtained images.

An essential step towards efficient use of image data is the image pre-processing before they are presented to the NN. Rather than taking raw image data as the inputs, the NN receives pre-processed data calculated from pairs of biased images, which we term a ``pseudo-PSF", as shown in Fig.~\ref{fig:MLAO_demonstration} and explained in the methods section. This pseudo-PSF contains information about the input aberration and is mostly independent of the unknown specimen structure. By removing the specimen information at this stage, we can reduce the demands on the subsequent NN, hence vastly simplifying the architecture required to retrieve the aberration information.

As most of the useful information related to aberrations was contained within the central pixels of the pseudo-PSF, a region of $32\times32$ pixels was extracted as the input to the NN. The first section of the NN was a bespoke convolutional layer that was designed to extract information from the inputs at different spatial scales.  The outputs from the convolutional layer were then provided to a fully connected layer, which was connected to the output layer. Full details of the NN design are provided in the methods and the supplementary document. This architecture -- rather unusually -- provided a link between the physical effects of aberrations on the imaging process and the mechanisms within the NN, specifically through the weights at the output of the first fully connected layer. 

NN training was performed using a diverse set of synthesised training data. These images were calculated using an appropriate model of the microscope imaging process in the presence of aberrations. Images were synthesised by convolutions of specimen structures with a PSF, incorporating various likely experimental uncertainties and noise sources. The specimens consisted of a range of artificial and realistic objects. Full details are provided in the methods.

This versatile concept could accommodate different aberration biasing strategies. Conventional modal sensorless AO methods typically required a minimum of $2N+1$ biased images to estimate $N$ aberration modes \cite{Facomprez2012}. However, the MLAO method has the ability to extract more information out of the images, such that aberrations could be estimated with as few as two images, although more biased images could provide better-conditioned information. In general, we defined methods that used $M$ differently biased images to estimate $N$ Zernike modes for aberration correction. The input layer of the NN was adjusted to accommodate the $M$ image inputs for each method. Out of the many possibilities, we chose to illustrate the performance using two biasing schemes: one using a single bias mode (astigmatism, Noll index \cite{Noll:s} $\text{i}=5$) and one using all $N$ modes that were being corrected. In the first case, we used either two or four images ($M=2$ or $4$) each with different astigmatism bias amplitude. We refer to these methods as \emph{ast2 MLAO} or \emph{ast4 MLAO}. Astigmatism was chosen as the most effective bias mode (see supplementary document, section 7). In the second case, biased images were obtained for all modes being estimated ($M = 2N$ or $4N$); this type is referred to in this paper as \emph{2N MLAO} or \emph{4N MLAO}. For a complete list of the settings for each demonstration, please refer to Table S2 in the supplemental document.

\section*{Results}\label{section:results}

In order to show its broad application, the MLAO method was demonstrated in three different forms of microscopy: 2-P and 3-P scanning microscopy and widefield 3-D SIM. This enabled testing in different applications to examine its performance coping with different realistic imaging scenarios. 

The MLAO methods were compared to two widely used conventional modal based sensorless AO methods (labelled as \emph{2N+1 conv} and \emph{3N conv}). The \emph{2N+1 conv} method used two biased images per modulation mode and an additional zero biased image to determine phase correction consisting $N$ modes simultaneously. The \emph{3N conv} method used three images per modulation mode (two biased and one unbiased images) and determined the coefficients of the modes sequentially. For both methods, the bias size was chosen to be $\pm 1$ rad for each mode. A suitable metric was selected to quantify the image quality. For each mode, the coefficients were optimised by maximising the quality metric of the corresponding images using a parabolic fitting algorithm. When used in 2-P and 3-P demonstrations, the total fluorescence intensity metric was optimised. For the widefield 3-D SIM microscope, a Fourier based metric was optimised \cite{Hall2020}. For the details of the two conventional methods, please refer to \cite{Booth2002,Facomprez2012}.

Different functions were defined as optimisation metrics for the conventional AO methods, and also to assist quantifiable comparisons of image quality improvement for the MLAO methods. These were defined as an intensity based metric $\text{y}_\text{I}$, a Fourier based metric $\text{y}_\text{F}$, a sharpness metric $\text{y}_\text{S}$ and a Fourier threshold based metric $\text{y}_\text{T}$. Details are provided in the methods section.

\subsection*{Two-photon microscopy}\label{section:results:2-P}

A range of method validations were performed on a 2-P microscope that incorporated a SLM as the adaptive correction element, including imaging bead ensembles and extended specimen structures. The experimental set-up of the 2-P system was included in Figure S9 (a) in the supplemental document. In order to obtain controlled and quantitative comparisons between different AO methods, the SLM was used to both introduce and correct aberrations. This enabled statistical analysis of MLAO performance with known input aberrations. System aberrations were first corrected using a beads sample before carrying out further experiments.

\begin{figure}
    \centering
    \includegraphics[width = 1.0\textwidth]{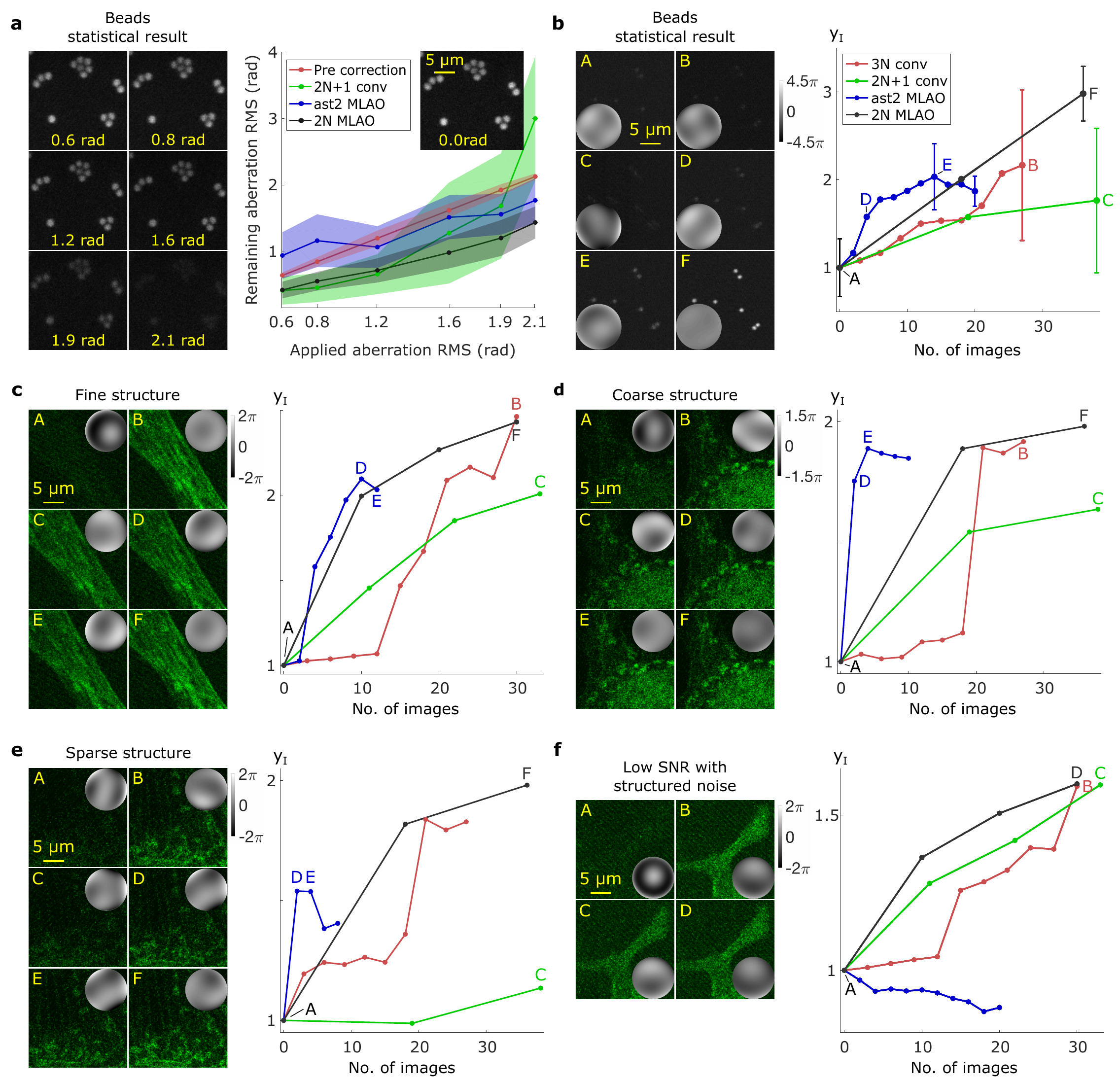}
    \caption{Comparative performance of MLAO methods in a 2-P microscope. (a) Residual aberration after one correction cycle for three methods. Points show the mean and the shaded area indicates the standard deviations (SDs) of aberration distributions.
    The images show an example field of view (FOV) when different amounts of a random aberration were introduced.
    (b)-(f) show the normalised intensity metric ($\text{y}_\text{I}$) as a proxy for correction quality, against the number of images used for multiple iterations of correction when random aberrations were introduced. In (b), an ensemble of ten random aberrations were corrected, imaging over the same FOV. Error bars on the plot showed the SDs of the fluorescence intensity before and after correction.
    (c)-(f) show specific corrections imaging microtubules of BPAE cells, illustrating performance for different specimen structures and imaging conditions. 
    The images were acquired before and after correction through the different methods (as marked on the metric plots). Insets on the images show residual wavefronts after correction for each image. The grayscale colorbars show phase in radians. }
    \label{fig:2-P_results_compile}
\end{figure}

We performed a statistical analysis to assess how MLAO algorithms (\emph{ast2 MLAO} and \emph{2N MLAO}) performed in various experimental conditions compared to conventional algorithms (\emph{2N+1 conv} and \emph{3N conv}). Experiments were conducted on fixed beads samples (Figure \ref{fig:2-P_results_compile} (a, b)), and Bovine Pulmonary Artery Endothelial (BPAE) cells (FluoCells$^\text{TM}$ Prepared Slide \#1) (Figure \ref{fig:2-P_results_compile} (c - f)). Dependent on the experiment, either $N=5$ or $N=9$ Zernike modes were estimated (see Table S2 in Supplemental document for details). 

\subsubsection*{Statistical performance analysis}

Figure \ref{fig:2-P_results_compile} (a) and (b) showed statistical comparisons of the different correction methods. Figure \ref{fig:2-P_results_compile} (a) displayed the residual aberrations gathered from twenty experiments, each consisting of one correction cycle from random initial aberrations including five Zernike modes. If the remaining aberration is below the pre-correction value, then the method provides effective aberration correction. A wide shaded area indicated inconsistent and less reliable correction. The results show that when correcting small aberrations with root mean square (RMS) = 0.63 to 1.19 rad, \emph{2N MLAO} performed similarly to \emph{2N+1 conv}. Between RMS = 1.19 to 1.92 rad, \emph{2N MLAO} corrected more accurately (lower mean aberration) and also more reliably (smaller error range). For large aberrations above RMS = 2.12 rad, \emph{2N+1 conv} completely failed, whereas the MLAO methods still improved aberration correction. \emph{ast2 MLAO} had poor performance at small aberrations (RMS = 0.63 to 0.84 rad) but provided reasonable correction for large aberrations (RMS = 1.92 to 2.12 rad). However, it is important to note that \emph{ast2 MLAO} required only two images for each correction cycle, far fewer that the ten and eleven images required respectively for \emph{2N MLAO} and \emph{2N+1 conv}.

Figure \ref{fig:2-P_results_compile} (b) displayed the mean value of metric $\text{y}_\text{I}$ from ten experiments against the number of images acquired during multiple iterations of the different correction methods. The corrected aberrations consisted of nine Zernike modes. It was shown that \emph{ast2 MLAO} corrects the fastest initially when the input aberration is large but converges to a moderate signal level, which indicates only partial correction of the aberration. \emph{2N MLAO} corrects more quickly and to a higher level than the conventional algorithms. The narrower error bars for both MLAO algorithms at the end of the correction process indicate that they are more reliable than the two conventional methods.


\subsubsection*{Correction on extended specimen structures}

Figure \ref{fig:2-P_results_compile} (c)-(f) showed experimental results when imaging microtubules of BPAE cells. Specimen regions were chosen to illustrate performance on different structures: (c) contained mainly aligned fine fibrous structures; (d) contained some large scale structures (bottom right); (e) contained fine and sparse features. For (f) we intentionally reduced illumination laser power and increased detector gain to simulate an imaging scenario with very low SNR. The images showed structured noise at the background, which could pose a challenge to estimation performance. A large randomly generated aberration (RMS = 2.12 to 2.23 rad) consisting of five (c and f) or nine (d and e) Zernike modes was used as the input aberration.

In (c), (d) and (e), \emph{ast2 MLAO} corrected the fastest initially when the aberration was large but converged to a moderate level of correction. \emph{2N MLAO} corrected faster in general than the conventional methods and converged to a higher level of correction. In (f) when SNR was poor and structured noise was present, \emph{ast2 MLAO} failed to correct while \emph{2N MLAO} continued to perform consistently.

\subsection*{Three-photon intravital microscopy}\label{section:results:3-P}
 
 Three-photon microscopy of neural tissue imaging is a particular challenge for sensorless AO, due to the inherently low fluorescence signal levels. While this could be alleviated by averaging over time, problems are created due to specimen motion. Further challenges are posed for functional imaging, due to the time dependence of emission from ion or voltage sensitive dyes. The demonstrations here show the robustness of the new MLAO methods in experimental scenarios where the conventional methods were not effective. Importantly, the MLAO methods were able to perform effective correction based on a small number of low SNR image frames without averaging.

\begin{figure}
  \centering
  \includegraphics[width = \textwidth]{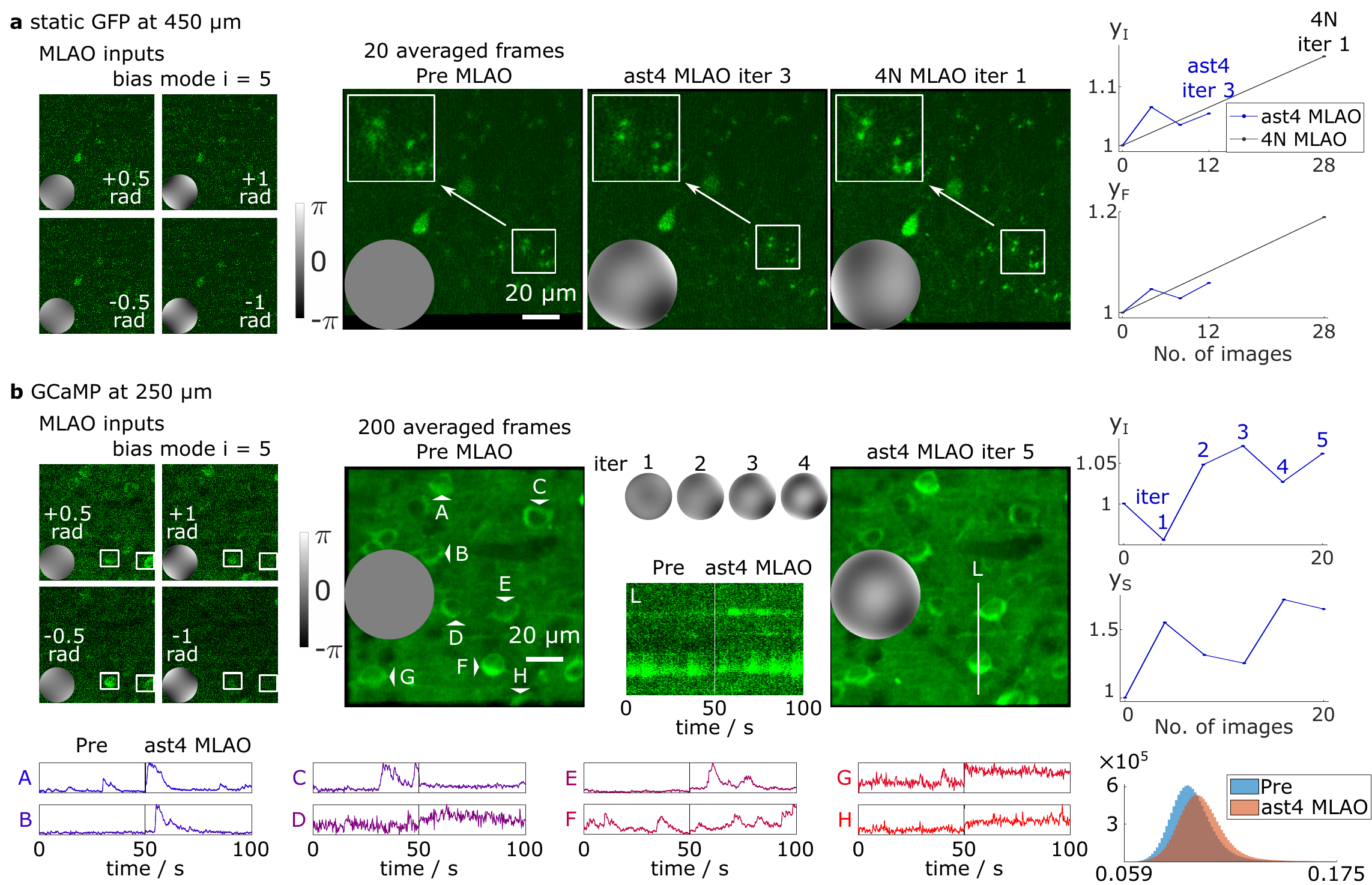}
  \caption{Aberration correction in three-photon microscopy of live mouse brains: (a) GFP-labelled cells at depth $450\mu m$ and (b) functional activity of GCaMP-labelled cells at $250\mu m$. Wavefronts inserted into the figures showed the phase modulations applied by the DM at the relevant steps; the common scale for each set of results is indicated by the grayscale bars in (a) and (b).  
  \\(a) shows on the left example single-frame images used in correction with the corresponding bias modes as insets; these were the image inputs to \emph{ast4 MLAO}. For \emph{4N MLAO}, six more bias modes and thus 24 more images were also used in each iteration.
  Three images at the central panel are shown averaged from 20 frames after motion correction. The rectangular boxes highlight regions of interest for comparison.
  The plots on the right show the intensity metric ($\text{y}_\text{I}$) and the Fourier metric ($\text{y}_\text{F}$), respectively, calculated from single image frames, against the number of images acquired for three correction iterations of \emph{ast4 MLAO} one correction iteration of \emph{4N MLAO}.\\
  (b) shows on the left example single-frame images used as inputs to the \emph{ast4 MLAO} correction with the corresponding bias modes as insets. White squares highlight two cells for comparison to show the fluorescence fluctuations over time due to neural activity.
  The central panel shows respectively before and after \emph{ast4 MLAO} correction through five iterations (iter 1 to 5), 200 frame averages after motion correction.
  The time traces were taken from the marked line L.
    The plots on the right show the intensity metric ($\text{y}_\text{I}$) and the sharpness metric ($\text{y}_\text{S}$), respectively, calculated from single image frames, against the number of images acquired for five iterations \emph{ast4 MLAO}.
    The lower panel shows the calcium activity of 8 cells (A-H marked on the averaged image). The lower right plot shows the histograms of the 200 frames collected before and after \emph{ast4 MLAO} corrections. The pixel values were normalised between 0 and 1.}
  \label{fig:3-P_results_compile}
\end{figure}

 The experimental set-up of the 3-P system is shown in Figure S9 (b) in the supplemental document. The microscope used an electromagnetic DM for aberration biasing and correction.  Two MLAO methods, \emph{ast4 MLAO} and \emph{4N MLAO}, were used to correct aberrations by using single frame images as inputs. In each case, more input frames were chosen than in the 2-P demonstrations, in order to cope with the lower SNR. The NNs were trained to estimate $N=7$ Zernike modes. 
 Two types of mice were used to perform live brain imaging of green fluorescent protein (GFP) labelled cells (Figure \ref{fig:3-P_results_compile} (a)) and functional imaging in GCaMP-expressing neurons (Figure \ref{fig:3-P_results_compile} (b)). In Figure \ref{fig:3-P_results_compile} (a), results were collected at $450\mu m$ depth and power at sample was 32 mW. In Figure \ref{fig:3-P_results_compile} (b), imaging was at $250\mu m$ depth and power at sample was 19 mW. Further 3-P results were included in the section 9 of supplemental document. For the details of the sample preparation, please refer to section 10B in supplemental document.
 

Figure \ref{fig:3-P_results_compile} (a) shows plots of the metrics $\text{y}_\text{I}$ and $\text{y}_\text{F}$ as proxies for correction quality when imaging  GFP labelled cells. Both \emph{ast4 MLAO} and \emph{4N MLAO} networks successfully improved the imaging quality. Similar to the \emph{ast2 MLAO} results in the 2-P demonstrations, \emph{ast4 MLAO} corrected more quickly at first, but converged to a lower correction level. In contrast, \emph{4N MLAO} performed better overall correction, but required more images. Panels ii-iv show averaged images in which blurry processes previously hidden below the noise level are revealed and get clearer through MLAO correction (as highlighted in the white rectangles). The example biased images shown in the left panel of Figure~\ref{fig:3-P_results_compile} (a) provide an indication of the low raw-data SNR that the MLAO method can successfully use.

Figure \ref{fig:3-P_results_compile} (b) shows results from imaging calcium activity in a live mouse. The \emph{ast4 MLAO} method successfully improved image quality despite the low SNR and fluorescence fluctuations of the sample. From both time traces of line 1 and cells A-H, it could be clearly seen that after corrections, signals were increase. The \emph{4N MLAO} method  failed to correct in this experimental scenario (results not shown). We will discuss the likely hypotheses for this in the discussion section. 

The fluctuating fluorescence levels due to neural activity mean that conventional metrics would not be effective in sensorless AO optimisation processes.  This is illustrated in Figure \ref{fig:3-P_results_compile} (b) iv and v, where it can be seen that no single metric can accurately reflect the image quality during the process of \emph{ast4 MLAO} correction. 
These observations illustrate the advantages of MLAO methods, as their optimisation process did not rely on any single scalar metric.

\subsection*{Widefield 3-D structured illumination microscopy}\label{section:results:widefieldDeepSIM}
The architecture of the NN was conceived so that it would be translatable to different forms of microscopy. In order to illustrate this versatility, and to complement to the previously shown 2-P and 3-P laser scanning systems, we applied MLAO to a widefield method. The 3D SIM microscope included multiple lasers and fluorescence detection channels and an electromagnetic DM as the correction element. Structured illumination patterns were introduced using a focal plane SLM. The detailed experimental set-up was included in Figure S9 (c) in the supplemental document. Further widefield results were included in the section 9 of supplemental document.

Without AO, 3D SIM reconstruction suffers artefacts caused by aberrations. Since typical specimens contain 3D structures, the lack of optical sectioning in widefield imaging means that the aberration correction process can be affected by out of focus light. As total intensity metrics are not suitable for conventional AO algorithms in widefield imaging, Fourier based sharpness metrics have often been used. However, such metrics depend on the frequency components of the specimen structure \cite{Antonello:20}. In particular, emission from out of focus planes can also affect the sensitivity and accuracy of correction. However, the NN based MLAO methods were designed and trained to mitigate against the effects of the sample structures and out of focus light. 

\begin{figure}
    \centering
    \includegraphics[width=0.65\textwidth]{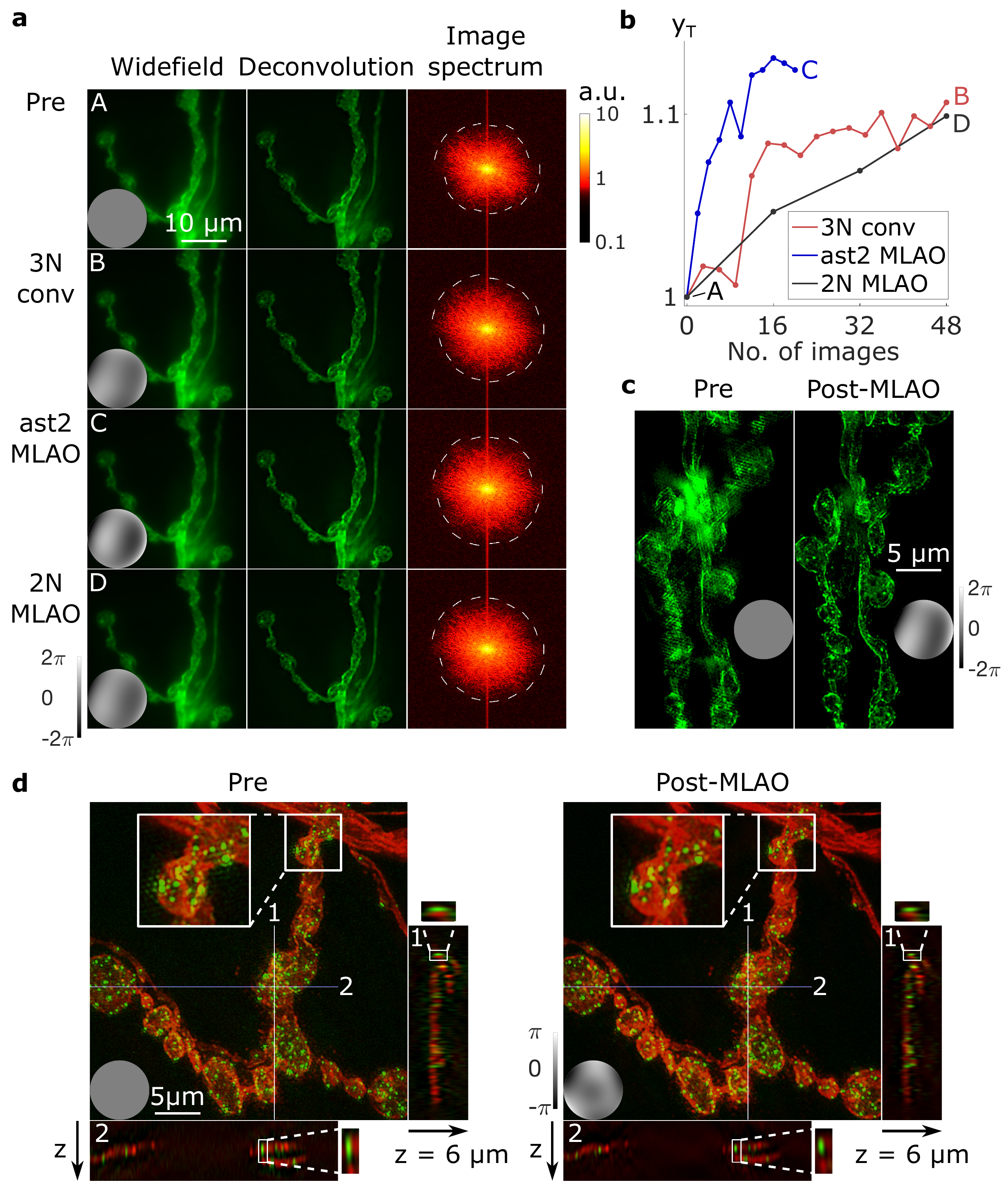}
    \caption{
    Aberration correction in a widefield 3-D structured illumination microscope (SIM).
    (a) Widefield images acquired A before and B-D after correction through different methods (as marked on the metric plot (b)). The second column shows corresponding deconvolved widefield images. 
The third column shows corresponding image spectra of the first column widefield images displayed in a logarithmic scale (as shown in the colorbar); dashed lines show the threshold where signal falls below the noise level.\\
    (b) The frequency threshold metric $\text{y}_\text{T}$ against the number of images, for two iterations of \emph{3N conv}, ten iterations of \emph{ast2 MLAO} and three iterations of \emph{2N MLAO}.
(c, d) 3-D projections of 3-D reconstructed SIM image stack of (c) $10\mu m$ and (d) $6 \mu m$ when by-passing AO and after five iterations of \emph{2N MLAO} correction. (d) Square inserts show zoomed-in region for comparison. x-z and y-z sections are shown through lines 1 and 2.\\
Insets to (a,c and d) show wavefronts corrected by the DM for each image acquisition; phase is shown on the adjacent scale bar.
}
    \label{fig:widefield_3D_SIM}
\end{figure}

Figure \ref{fig:widefield_3D_SIM} shows results from two NN-based methods \emph{ast2 MLAO} and \emph{2N MLAO} compared to the conventional algorithm \emph{3N conv}, which used the $\text{y}_\text{S}$ metric.  Sensorless AO was implemented using widefield images as the input (Figure \ref{fig:widefield_3D_SIM} (a, b)). The correction settings thus obtained by the \emph{2N MLAO} method were then applied to super-resolution 3D SIM operation (Figure \ref{fig:widefield_3D_SIM} (c, d)). $N=8$ Zernike modes were involved in the aberration determination. The specimen was a multiple labelled \textit{Drosophila} larval neuromuscular junction (NMJ). For the details of the sample preparation, please refer to section 10B in supplemental document.

Figure \ref{fig:widefield_3D_SIM} (b) showed that \emph{ast2 MLAO} corrected most quickly; \emph{2N MLAO} corrected similarly as \emph{3N MLAO} and were less effective.
Figure \ref{fig:widefield_3D_SIM} (a) showed the effectiveness of correction on raw and deconvolved widefield images. The third column showed the changes in image spectrum of the widefield images after correction. The dashed line shows a threshold where signal falls below the noise level. It can be seen that all three methods increased high frequency content compared to (A) before AO correction. Figure \ref{fig:widefield_3D_SIM} (c) and (d) showed the images after 3D SIM reconstruction. It can be clearly seen that when by-passing AO (left hand side), there were strong artefacts due to aberrations. After correcting using five iterations of \emph{2N MLAO}, artefacts were suppressed and z-resolution was improved (see sections through line 1 and 2 in Figure \ref{fig:widefield_3D_SIM} (d)) 

\section*{Discussion}\label{Sec:discussion}

The power and simplicity of the MLAO method arise mainly from a combination of three aspects: the pre-processing of image data, the bespoke NN architecture, and the definition of the training data set. All of these aspects are informed by physical and mathematical principles of image formation. This forms a contrast with many other data-driven deep learning approaches, where complex NNs are trained using vast amount of acquired data.

The calculation of the pseudo-PSF from pair of biased images (as shown in Figure \ref{fig:MLAO_demonstration} (c) and elaborated in the Methods) acts to remove most of the effects of unknown specimen structure from the input data. The information contained within the pseudo-PSF encodes indirectly how aberrations affect the imaging PSF (see Figure S2 in the supplemental document for more details). There is a spatial correspondence between a pixel in the pseudo-PSF and the PSF itself. Hence, spatial correlations across the pseudo-PSF relate to spatial effects of aberrations on the images. 

The set of pseudo-PSFs forms the input to the convolutional layers of the NN. The masks in each convolutional layer probe, in effect, different scales across the pseudo-PSF.  Hence, one can attribute a correspondence between the output of these layers and the effects aberrations have over different physical scales in the image. Such phenomena are heuristically demonstrated in section 3 of the supplementary information. By extracting relevant weight connections from inside the NN, we can observe embedded physical interpretations of how the machine learned to process aberration information contained in images. 

To illustrate this, we extracted from the trained NN the weights between the layer embedding physical interpretations and the next fully connected layer (marked by the red arrows in Figure \ref{fig:MLAO_demonstration} (a) and the red arrow enclosed by the dashed square in Figure S1 in the supplemental document). Going down the convolutional layers, the scale of probed features increases from a single pixel, through small scale features, up to large scale features (as explained in section 4 of the supplemental document). The RMS values of the weights from each convolutional layer are shown in Table \ref{tab:weight-analysis}, where the data are shown for the ensembles of the two classes of MLAO networks used in this paper, \emph{ast$X$ MLAO} and \emph{$X$N MLAO} (where $X=$2 or 4). A full breakdown is provided in the Figure S4 of the supplementary document.

\begin{table}
    \centering
    \begin{tabular}{|c|c|c|c|c|c|}
    \hline
         Layer&1&2&3&4&5\\
         \hline
         \emph{ast$X$ MLAO}&0.23&0.19&0.17&0.18&0.23\\
         \hline
         \emph{$X$N MLAO}&0.39&0.14&0.15&0.13&0.20\\
         \hline
    \end{tabular}
    \caption{The RMS of the weight distributions extracted from different convolutional layers of the two classes of trained CNNs, \emph{ast$X$ MLAO} and \emph{$X$N MLAO}. The values shown are calculated from the ensemble of corresponding layers from all CNNs of the given class.}
    \label{tab:weight-analysis}
\end{table}

The largest weight variation was in the first layer in the \emph{$X$N MLAO} NN, which indicates that this algorithm extracts more information from the single pixel detail than from larger scale correlations. In contrast, \emph{ast$X$ MLAO} assigns weights more evenly across all layers. As explained in the supplementary document, the single pixel extraction from the pseudo-PSF is related to the Strehl ratio of the PSF and the intensity information of the images in non-linear systems. Hence, it is expected that the \emph{$X$N MLAO} NN, which uses as similar set of bias aberrations to the conventional method, would learn as part of its operation similar behaviour to the conventional algorithm. The same phenomena can also explain why in 3-P GCaMP imaging of neural activity \emph{ast$X$ MLAO} was less affected by the fluorescence fluctuations than \emph{$X$N MLAO}, as \emph{ast$X$ MLAO} relies less on overall fluorescence intensity changes. Similarly, in widefield imaging \emph{ast$X$ MLAO} was more effective at extracting PSF variations than \emph{$X$N MLAO} as the overall fluorescence intensity did not change with aberrations in single-photon imaging. Conversely, \emph{ast$X$ MLAO} generally performed worse than \emph{$X$N MLAO} in 2-P imaging when structured noise present, as \emph{ast$X$ MLAO} used fewer images and hence had access to less detectable intensity variations than \emph{$X$N MLAO}. The fact that \emph{ast$X$ MLAO} had access to less well-conditioned image information may also explain why in general it was able to correct aberrations to a lower final level than \emph{$X$N MLAO}.

\section*{Conclusion}

The MLAO methods achieved the aims explained at the outset. They provided more efficient aberration correction with fewer images over a larger range, reducing time required and specimen exposure. The training procedure, which was based on synthesised data, ensured that the AO correction was robust to uncertainty in microscope properties, the presence of noise, and variations in specimen structure. The concept was translatable across different microscope modalities, simply requiring training using a revised imaging model.

The new methods used NN architectures that are orders of magnitude simpler, in terms of trainable parameters, than in previous similar work (see supplementary information, section 6). This vast simplification was achieved through pre-processing of data to remove most of the effects of unknown specimen structure. The physics-informed design of the NN also meant that -- unusually for most NN applications -- the learned weights inside the network provided indications of the physical information used by the network. This provides constructive feedback that can inform future AO system designs and the basis for extension of the MLAO concept to more demanding tasks in microscopy and other imaging applications.

\section*{Methods}


\subsection*{Image pre-processing}\label{section:post_image_processing}

Image data were pre-processed before being used by the NN, in order to remove effects of the unknown specimen structure. The resulting ``pseudo-PSFs'' were better conditioned for the extraction of aberration information, independently of the specimen. The image formation can be modelled as a convolution between specimen fluorescence distribution and an intensity PSF. The AO introduced pre-chosen bias aberrations, so that multiple images with different PSFs could be acquired over the same FOV. Mathematically, this process can be expressed as 
\begin{align}
 I_1 = O\ast f_1+\delta_1\nonumber \\ 
 I_2 = O\ast f_2+\delta_2\label{eq:image_formation}
\end{align}
where $I_1$ and $I_2$ were the images acquired with two different PSFs $f_1$ and $f_2$ for the same unknown specimen structure $O$. $\delta_1$ and $\delta_2$ represent combined background and noise in each image. In order to remove (or at least reduce) the effects of specimen structures, we defined the pseudo-PSF as
\begin{align*}
 \text{pseudo-PSF} = \mathcal{F}^{-1}\left[\frac{\mathcal{F}(I_1)}{\mathcal{F}(I_2)}\right] &= \mathcal{F}^{-1}\left[\frac{\mathcal{F}(O\ast f_1+\delta_1)}{\mathcal{F}(O\ast f_2+\delta_2)}\right]\\ 
 &= \mathcal{F}^{-1}\left[\frac{\mathcal{F}(O)\times \mathcal{F}(f_1)+\mathcal{F}(\delta_1)}{\mathcal{F}(O)\times \mathcal{F}(f_2)+\mathcal{F}(\delta_2)}\right]
\end{align*}
where $\mathcal{F}$ was the 2D Fourier transform and $\mathcal{F}^{-1}$ was its inverse (see Figure \ref{fig:MLAO_demonstration} (c)). The term ``pseudo-PSF'' was chosen as the function was defined in the same variable space as a PSF, although it is not used directly in any imaging process.  A similar computational process was shown elsewhere for different applications using defocussed images \cite{Xin:19}. Assuming the noise is small enough to be neglected
\begin{equation}
\text{pseudo-PSF} = \mathcal{F}^{-1}\left[\frac{\mathcal{F}(I_1)}{\mathcal{F}(I_2)}\right]\approx \mathcal{F}^{-1}\left[\frac{\mathcal{F}(f_1)}{\mathcal{F}(f_2)}\right] 
\end{equation}
There is an implicit assumption here that there are no zeroes in the object spectrum $\mathcal{F}(O)$ or the optical transfer function $\mathcal{F}(f_2)$. In practice, it was found that a small non-zero value of $\mathcal{F}(\delta_2)$ mitigated against any problems caused by this. Furthermore, although structured noise was present in the pseudo-PSFs (see e.g. Figure S1 in the supplemental document), it was found that this did not detrimentally affect data extraction through the subsequent NN.  As a further mitigation, we calculated pairs of pseudo-PSFs from pairs of biased input images by swapping the order from $(f_1, f_2)$ for the first pseudo-PSF to $(f_2, f_1)$ for the second.

Example pseudo-PSFs are shown in  Figure S1 and S2 in the Supplemental document. As most information was contained within the central region, to ensure more efficient computation, we cropped the central region ($32\times32$ pixels) of the pseudo-PSFs to be used as the input to the NN. Dependent upon the MLAO algorithm, the input to the NN would consist of a single pair of cropped pseudo-PSFs, or multiple pairs corresponding to the multiple pairs of bias aberrations applied in different modes.

\subsection*{Neural network training}\label{sec:neural_network_training}
To estimate phase aberrations from pseudo-PSFs, a convolutional based neural network was designed incorporating physical understanding of the imaging process and was trained through supervised learning. Synthetic data were used for training and the trained networks were then tested on real AO microscopes. For each imaging modality (i.e. 2-P, 3-P and widefield), a separate training dataset was generated, with the imaging model and parameters adjusted for different applications. For the details of neural network architecture and synthetic training data generation, please see section 1 and 2 of suplementary information.

\subsection*{Image quality metrics}

Different image quality metrics were defined for use as the basis for optimisation in conventional sensorless AO methods and as proxies to quantify the level of aberration correction.
$\text{y}_\text{I}$ is an intensity based metric and can be used in non-linear imaging systems. It is defined as 
$$\text{y}_\text{I}=\sum_{i=1}^{l} T(i)$$
where $T(i)$ is a flattened array of image $I(x)$ after sorting pixel values in descending order (indexed i). $\text{y}_\text{I}$ is computed to sum only the first $l$ pixel values to provide a fair quantitative intensity variation analysis when imaging sparse samples. $l$ was adjusted for different experiments depending on the density of the sample structures and was chosen to be always larger than 200.

$\text{y}_\text{F}$ is a Fourier based metric and provides an alternative aspect to the intensity metric. It is defined as 
$$\text{y}_\text{F}={\int\int}_{0.1f_{max} < |f| < 0.6f_{max}}|\mathcal{F}[I(x)]| d^2 f$$ 
where $\mathcal{F}[I(x)]$ is the 2D Fourier transform of image $I(x)$ from $x$ domain to $f$ domain; $f_{max}$ is the maximum frequency limit of the imaging system. The range $0.1f_{max}<|f|<0.6f_{max}$ was selected such that most PSF related frequency information was included in the range.

$\text{y}_\text{S}$ is a sharpness metric that can be used for optimisation in widefield systems, where the other metrics are not practical, or applications with fluorescence fluctuations. It is defined as $$\text{y}_\text{S}=\frac{ {\int\int}_{nf_{max}<|f|<mf_{max}}|\mathcal{F}[I(x)]| d^2 f}{{\int\int}_{0<|f|\leq nf_{max}}|\mathcal{F}[I(x)]| d^2 f}$$ where $1>m>n>0$. This metric is defined as the ratio of higher to lower spatial frequency content, which is dependent upon aberration content, but independent of changes in overall brightness. $m$ and $n$ can be adjusted for different imaging sample structures such that the frequency components $0<|f|<nf_{max}$ contain mainly sample features and $nf_{max}<|f|<mf_{max}$ captures mainly PSF sharpness. In figure \ref{fig:3-P_results_compile} (b), $n$ was chosen to be $0.05$ and $m$ was chosen to be $0.6$.

$\text{y}_\text{T}$ is a frequency threshold metric that can be used to analyse the image quality in widefield systems. It is defined as
$$\text{y}_\text{T}=\frac{\int_0^{2\pi} \,\,f_T(\theta)\,d\theta}{\int_0^{2\pi} d\theta}$$
where $f_T(\theta)$ is the maximum frequency component for each angular segment $\theta$ such that $ |FI( \,\,\forall \, f<f_T(\theta),\theta)| \geqslant T$. $FI(f,\theta) = \mathcal{F}[I(x)]$ is the Fourier transform of image $I(x)$ expressed in polar coordinates $(f,\theta)$. $T$ is a threshold value such that $|FI(f,\theta)| < T$ can be considered as noise. 

\subsection*{Microscope implementations}
Three microscopes were used to demonstrate and examine the MLAO method. The microscope implementations are briefly described here and fully elaborated in the supplementary document section 10A.

In the home built 2-P system, a Newport-Spectra-Physics DeepSee femtosecond laser was used as the illumination with wavelength set at $850 nm$. Light was modulated by a Hamamatsu spatial light modulator before passing through a water immersion objective lens with NA equals to 1.15 and reaching the sample plane.

A commercial Scientifica microscope system was used as the basis for our 3-P demonstration. In the 3-P system, a femtosecond laser passed through a pair of compressors and operated at $1300 nm$. Light was modulated by a Mirao 52E deformable mirror before reaching a water dipping objective lens with NA equals to 0.8.

In the home built widefield 3D SIM system, two continuous wave lasers with wavelengths equal to 488 and $561 nm$ were used as the illumination. Light was modulated by a ALPAO 69 deformable mirror before reaching a water dipping objective lens with NA of 1.1.

\subsection*{Image acquisition and processing}

For 3-P imaging of live specimens, where motion was present, averaging was performed after inter-frame motion correction using TurboReg \cite{turboreg}. Time traces were taken from 200 raw frames captured at 4 Hz consecutively for each of the pre- and post-MLAO corrections.

For the widefield/SIM results, widefield images were processed where indicated using the Fiji iterative deconvolution 3-D plugin \cite{FijiDeconvolve2022}. A PSF for deconvolution was first generated using the Fiji plugin Diffraction PSF 3-D with settings the same as the widefield microscope. For the deconvolution, the following settings were applied: Wiener filter gamma equals to 0; both x-y and z direction low pass filter pixels equal to 1; maximum number of iterations equals to 100; and the iteration terminates when mean delta is smaller than 0.01\%. 

The thresholds shown on the widefield image spectra were calculated by identifying the largest frequency in all x-y directions with image spectrum components higher than noise level. The noise level was identified by averaging the components of the high spectral frequency, i.e. at the four corners of the image spectrum. Starting from the lowest frequency, each angular and radial fragment was averaged and compared to the noise level. The largest component which was still above the noise level was traced on the image spectra by the dashed line and identified as the threshold.

Each 3D-SIM frame were extracted from a set of 15 image frames using the SoftWorx package (Applied Precision). \cite{GUSTAFSSON20084957} The projected images were obtained by summing frames at different z depths into an extended focus xy image.

\bibliography{sample}



\section*{Acknowledgements}

This work was supported by grants from the European Research Council (to MJB: AdOMiS, No. 695140, to AMP: No. 852765), Wellcome Trust (to MJB: 203285/C/16/Z, to ID and MJB: 107457/Z/15/Z, to AMP: 204651/Z/16/Z, to HA: 222807/Z/21/Z), Engineering and Physical Sciences Research Council (to MJB: EP/W024047/1).

\section*{Author contributions}

QH and MJB conceived the overall physics-informed approach including data pre-processing and bespoke NN architecture. MH, QH and MJB developed NN architectures and the training approach. QH, MH, MW, JA and DS developed the software packages. JW, QH, AMP set up the microscopes for the experimental demonstrations.
QH performed the two-photon experiments, supervised by MJB.
HA, JW and QH performed the three-photon experiments, supervised by AMP and MJB.
JW, MW, DS, QH and RMP performed the widefield/SIM experiments, for which DG, TC and RMP prepared specimens, supervised by ID and MJB.
QH performed data analysis. QH and MJB wrote the manuscript. All authors reviewed the manuscript.

\section*{Data availability statement}
The datasets generated during and/or analysed during the current study are available from the corresponding author on reasonable request.

\section*{Additional information}

All experimental procedures involving animals were conducted in accordance with the UK animals in Scientific Procedures Act (1986).





\end{document}


\maketitle
\section{MLAO process and CNN architecture}\label{Sec:supplimental:CNN_architecture}
The MLAO aberration estimation process consists of two parts: image pre-processing to compute pseudo-PSFs from images and a CNN-based machine learning process for mode coefficient determination. 
A stack of M images over the same field of view, each with a different pre-determined bias phase modulation, was used to calculate pseudo-PSFs according to the procedure in the methods section. It was observed and understood that most of the information was contained within the central region of the calculated pseudo-PSFs.
\footnote{The process of calculating pseudo-PSFs can be interpreted as a deconvolution between two PSFs. Depending on the sampling size of the imaging system, most details of a deformed PSF typically occupy a central region of a few pixels. Most features of the pseudo-PSFs were thus captured within the central region.} 
A central patch of $32\times 32$ pixels was then cropped and used as the inputs to the CNN. Cropped pseudo-PSFs were processed by a sequence of convolutional layers (CL) with trainable $3\times 3$ kernels, each followed by a local $2\times 2$ max-pooling and thus the x and y sizes were reduced by half but the stack size was increased twice going down each CL. For the input pseudo-PSFs and each of the CL outputs, a global max-pooling was applied and concatenated into a fully connected layer (FCL). This concatenated FCL was connected to the next FCL containing 32 neurons, which in turn was connected to the output layer, which produced the coefficients of the N chosen Zernike modes. The activation functions were chosen to be tanh and linear (only for the last layer connection FCL 32 and the output). The regularizer used was L1L2, the initializer was glorot-uniform and the optimizer was AdamW. The CNN architecture was built and the network training was conducted using TensorFlow.\cite{tensorflow2015-whitepaper} As elaborated in the results section of the manuscript, M and N may be varied to suit different applications.

The weights in the connection between the concatenated FCL and FCL32 (enclosed by a grey dashed square) were extracted and analysed to understand the physical significance of structures in the pseudo-PSFs in influencing the learning of the CNN. Further analysis of such weights is provided in Discussion of the main paper and section \ref{Sup:Sec:weight-analysis} of this document.

\begin{figure}[H]
    \centering
    \includegraphics[width=1.0\textwidth]{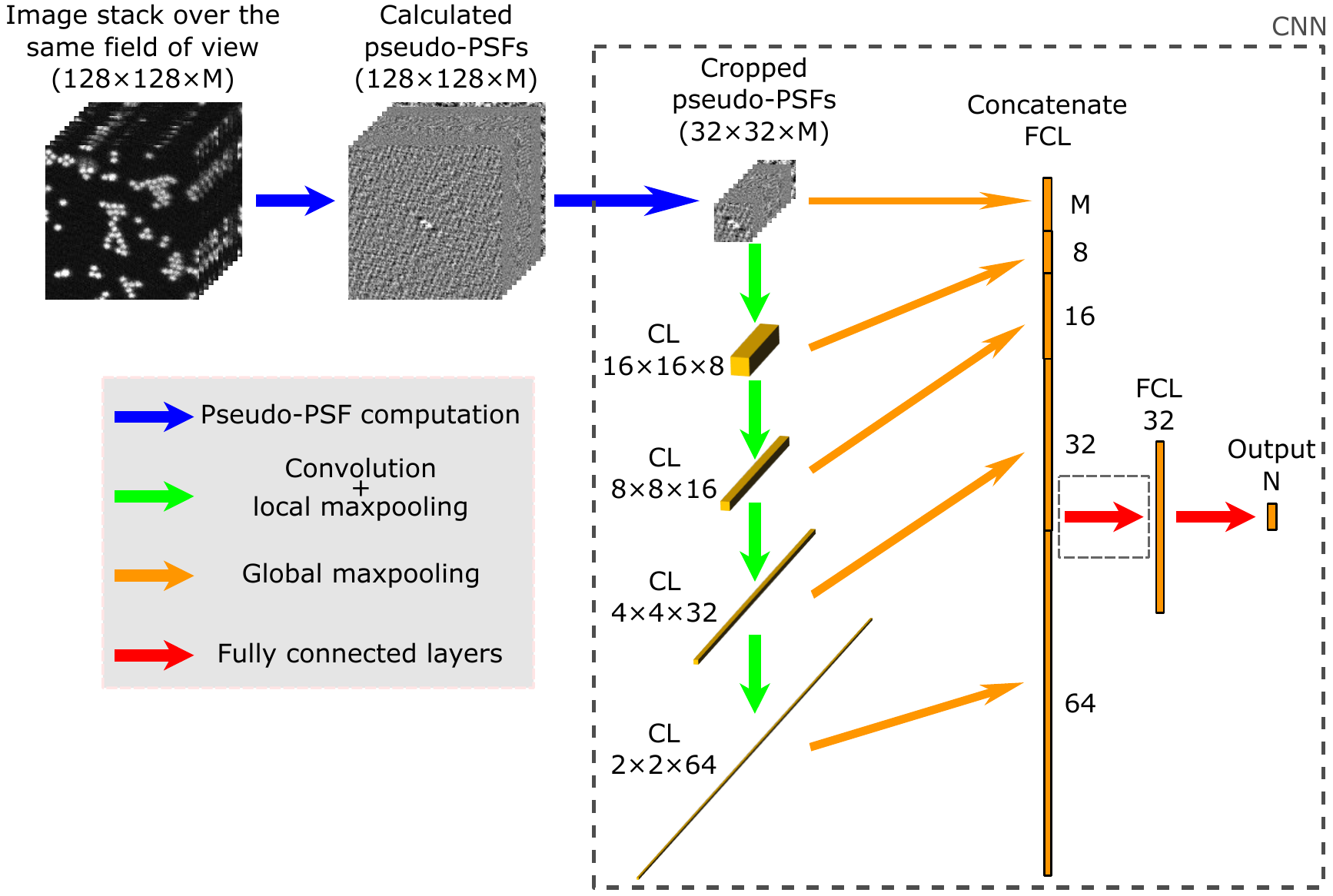}
    \caption{A schematic illustration of the MLAO process and CNN architecture (enclosed by a black dashed square) designed for phase determination applications. CL: convolutional layer followed by local max-pooling; FCL: fully connected layer; M: number of input images and computed pseudo-PSFs; N: number of estimated output Zernike modes.}
    \label{fig:network_structure}
\end{figure}

\section{Training data synthesis}

Due to the impracticality of acquiring sufficient high-quality data experimentally, a large dataset of simulated image data was constructed. The simulations were designed to resemble images collected from different microscopes when imaging a range of samples.

We started with a collection of image stacks (containing around a total of 350 images) obtained from high-resolution 3D microscopy of various specimens labelled with nuclear, cytoplasmic membrane and/or single-molecule markers. The images were down-sampled to 8-bit (128$\times$128) and separated into their individual channels. This formed a pool of realistic sample structures which were later used to generate synthetic images. To further augment the varieties of sample structures, random rotations were applied and synthetic shapes including dots, rings, circular shapes, curved and straight lines of varying sizes were randomly introduced.

The simulated training dataset was generated by convolving the sample structures with synthetic PSFs, $f$ (see Eq. (1) in the main paper
). $f$ was modelled as a pixel array through
\begin{equation}
    f = \left|{\mathcal{F}\left(P \,e^{j \left({ \Psi + \Phi + \Xi }\right) }\right)}\right|^l\label{eq:psf_generation}
\end{equation}
where $\mathcal{F}$ represented the 2D discrete Fourier transform. $P$ was the circular pupil function, defined such that pixels in the region outside the pupil had value zero. The ratio between the radius of the pupil in pixels and the size in pixels of the overall array was adjusted to match sampling rates for different microscopes. 
In practical scanning optical microscopes, the sampling rates can be easily adjusted, although perhaps not arbitrarily. Hence, for experimental flexibility, the ratio for the simulated training dataset was tuned to be within the range of $1.0\times$ to $1.2\times$ the base sampling rate. The base sampling rate was defined as using two pixels to sample the full width half maximum (FWHM) of the PSF of the system when aberration free. For the widefield system, the ratio was tuned to simulate the projection of the camera pixel sampling rate at the specimen.
Figure S5 in the supplemental document shows how tolerable a trained network was when tested on data collected at different pixel sampling. $P$ also incorporated the illumination profile for different practical imaging systems, such as when using truncated Gaussian illumination at the pupil in the 3-P microscope. The exponent $l$ varied with imaging modes: when simulating a 3-P, a 2-P and a single photon widefield microscope, $l$ was set to 6, 4 and 2 respectively. 

The total aberration was expressed as a sum of chosen Zernike polynomial modes $\Psi + \Phi + \Xi = \sum_i a_i Z_i$.  $\Psi$ was the sum of the randomly generated specimen aberrations, which included all modes that the AO system was designed to correct. $\Phi$ represented the additional bias aberrations. $\Xi$ included additional non-correctable higher order Zernike modes. The coefficients of the correctable modes were randomly generated for each data set. Representing the set of coefficients $\left\lbrace a_i \right\rbrace$ as a vector $\mathbf{a}$, the random coefficients followed a modified uniform n-sphere distribution \cite{marsaglia1972} where both the direction and the two-norm of  $\mathbf{a}$ were uniformly distributed. The maximum two-norm (size) of $\mathbf{a}$ were chosen differently for different imaging applications. This distribution allowed a denser population close to zero aberration, which was intuitively beneficial to train a stable NN. We also added random small errors to the correctable coefficients so that the labels were slightly inaccurate. This was to simulate situations when the AO would be incapable of introducing perfect Zernike modes. The spurious high order non-correctable Zernike modes were included to further resemble realistic scenarios in a practical microscope.

Poisson, Gaussian, pink and structured noise of varying noise level were also introduced to the generated images after the convolution to allow the training dataset to simulate more closely real microscope images.

Note that the scalar Fourier approximation of Eq. \ref{eq:psf_generation} was chosen for simplicity, although more accurate, vectorial, high numerical aperture (NA) objective lens models could have been applied \cite{Ignatowski1919,Richards1959,StamnesJakobJ1986Wifr,BORUAH20094660}. Although the chosen model would deviate from high NA and vectorial effects, the main phenomena under consideration here -- namely the effects of phase aberrations on PSFs and images -- are adequately modelled by scalar theory.

\section{Zernike polynomials and example pseudo-PSFs}\label{Sec:supplimental:Zernike-pseudo-PSF}
\begin{figure}
    \centering
    \includegraphics[width=\textwidth]{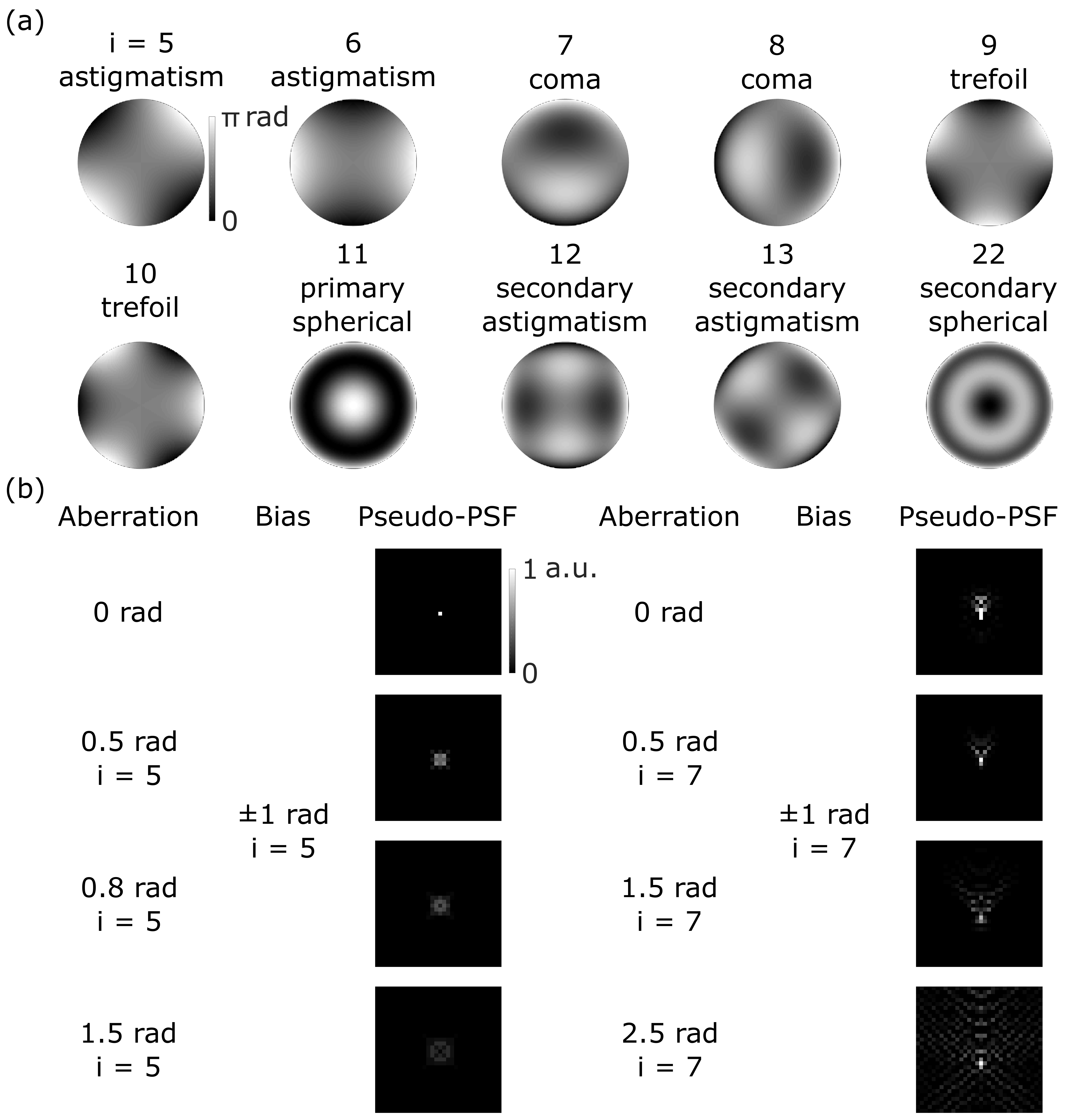}
    \caption{(a) Zernike polynomials Noll's index 5-13, 22. This is a whole list of the polynomials used for aberration determinations in the paper. (b) Examples of pseudo-PSFs. The first column is the input aberration and the second column is the bias mode used in pseudo-PSFs generation.}
    \label{fig:Zernike-pseudo-PSF}
\end{figure}

A total of ten Zernike polynomials were used for aberration estimation and correction presented in the paper. A list of the polynomials, sequenced using Noll's indices, were included in Figure \ref{fig:Zernike-pseudo-PSF} (a).

Figure \ref{fig:Zernike-pseudo-PSF} (b) included some examples of pseudo-PSFs. It can be observed that when aberration size increases, the maximum pixel value of the Pseudo-PSF decreases; a global max-pooling of the pseudo-PSF extracts information related to the Strehl ratio of the PSFs. Pseudo-PSFs also have shapes that are related to the aberrated PSF shapes.

\section{Physical information embedded in the CNN architecture }\label{Sec:supplimental:physical_meaning}
As mentioned in the main paper, the bespoke CNN architecture embedded information about the physical effects of aberrations on images within the trainable parameters.
 To illustrate these phenomena, we designed six input patterns and two filters to calculate how values obtained after global max-poolings from different convolutional layers were related to the features of the patterns. Normally, the filters would be learned as part of the training process, but for illustrative purposes, we have defined them manually here.

\begin{figure}
    \centering
    \includegraphics[width = \textwidth]{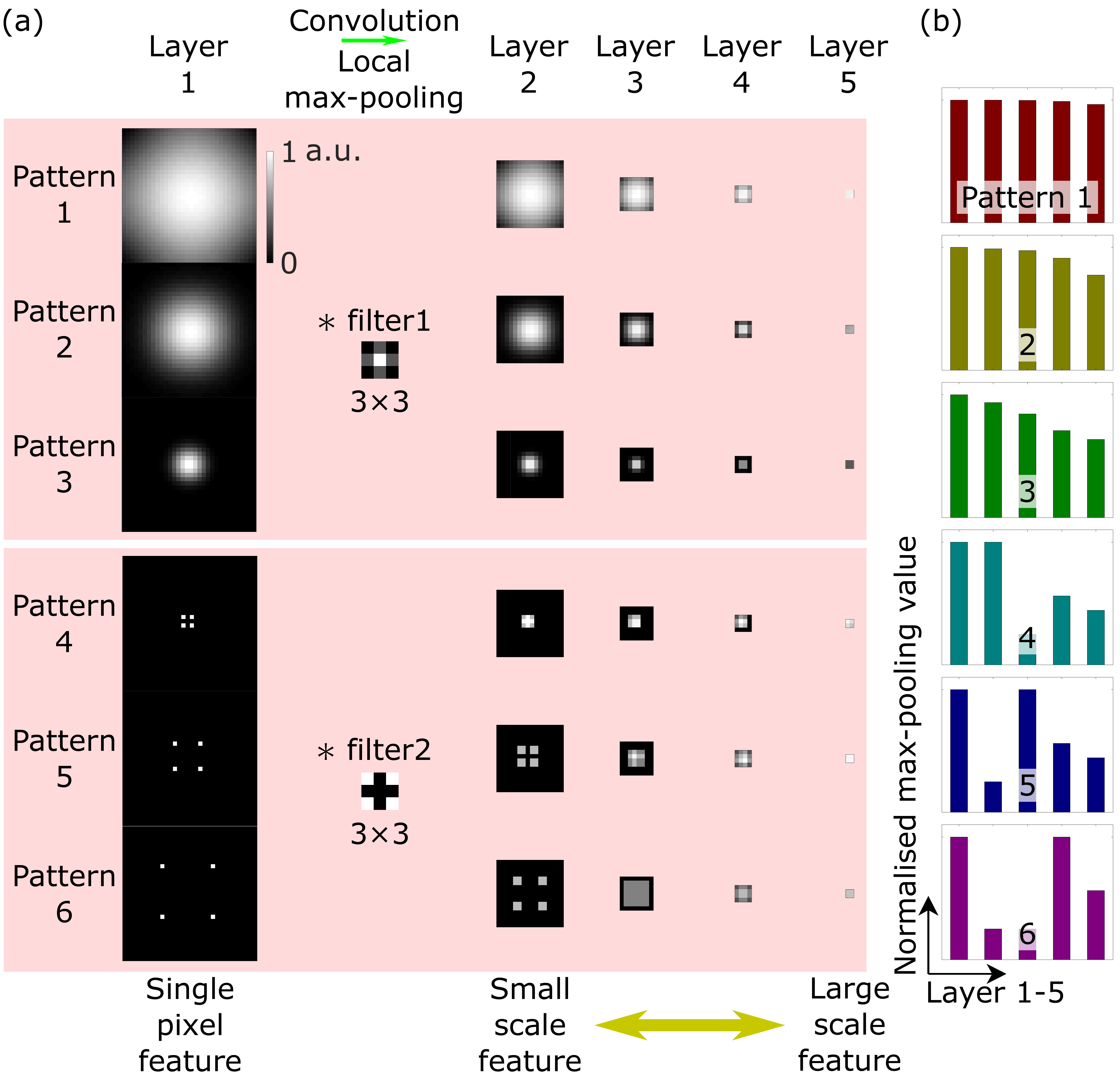}
    \caption{Demonstrations of the link between feature sizes and convolutional layers. (a) Pattern 1 to 6 each underwent a series of convolutions followed by a $2\times 2$ local max-pooling. Pattern 1 to 3 were convolved with filter 1 and pattern 4 to 6 were convolved with filter 2. For each layer, a global max-pooling were carried out to extract the maximum reading of each layer. The physical interpretations of the extracted values of the different layers were related to Strehl ratio (layer 1) and shapes with features ranging from small scales (layer 2) to large scales (layer 5). The extracted readings was normalised with the readings of their respective previous layer and displayed in (b). The horizontal axis of each plot in (b) indicates from which layer the normalised maximum reading (indicated by the vertical axis) was extracted from.}
    \label{fig:CNN_architecture_physical_meaning}
\end{figure}

As shown in Figure \ref{fig:CNN_architecture_physical_meaning}, patterns 1 to 3 had the same general shape but varying sizes. They were all convolved with the same filter 1. Pattern 1 had the largest feature and the values obtained were almost constant throughout layers 1 to 5 (see Figure \ref{fig:CNN_architecture_physical_meaning} (b)). Patterns 2 and 3 had smaller features and the extracted values reduced when moving further down the layers, where the embedded physical scales were more closely related to large scale features. Patterns 4 to 6 had the same general shape with four peaks positioned at the corners of a square. They were all convolved with filter 2, which shared a similar general shape. Pattern 4 had the smallest feature size and resulted a largest value in layer 2. Patterns 5 and 6 had larger feature sizes and resulted in largest values in layers 3 and 4, respectively. This trend confirms the expectation that layers later in the CNN probe larger scales in the input images.  Note that all the patterns were designed in such a way that the maximum pixel reading (and thus the value max-pooled from layer 1) equalled to 1.

\section{Weight analysis of different trained neural networks}\label{Sup:Sec:weight-analysis}

\begin{figure}
    \centering
    \includegraphics[width = 0.7\textwidth]{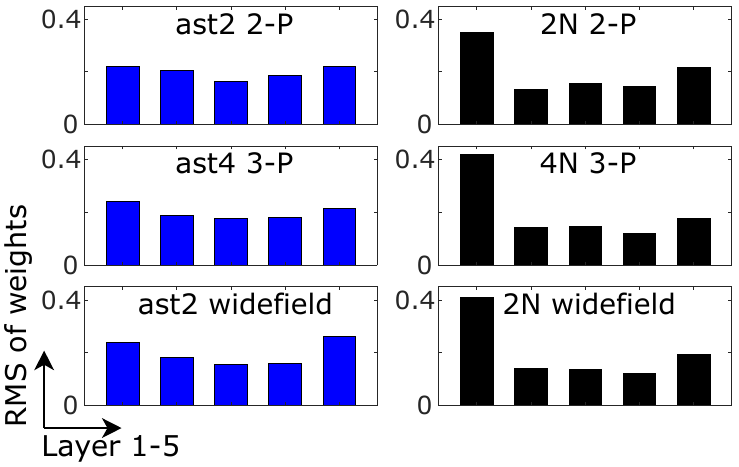}
    \caption{Analysis of the weight distributions across convolutional layers in the CNNs trained for different biasing schemes and microscopes. }
    \label{fig:layer_weights}
\end{figure}

Figure~\ref{fig:layer_weights} shows the root-mean-square (RMS) values of the weights at the output of each section of the concatenated FCL following the convolutional layers of the CNN.  These weights encode information about  physical phenomena in the pseudo-PSF that is related to the spatial effects of aberrations on images. Higher numbered layers correspond to larger scale features. Similar distributions are seen for all of the \emph{ast} CNNs class and all of the \emph{2/4N} class. Most notably, it can be seen that the \emph{2/4N} networks all carry heavier weights in layer 1, which is most similar to the Strehl ratio variations of the PSFs.

\section{Trainable neural network parameters}
The bespoke NN and data pre-processing steps were designed with knowledge of the physical basis of image formation. This permitted signficant reduction in NN complexity compared to previous methods for aberration estimation.  This architecture not only allowed improved performances, providing insights on internal workings, but also had a structure size orders of magnitude smaller than common NNs used in similar applications (see the comparison in Table \ref{tab:NN_parameter_size_comparison}). This will be beneficial for future applications as NN with fewer trainable parameters would generally require less training data and a shorter training time. Furthermore, the simplified design means that there is greater potential for extending the method to more challenging applications.
\begin{table}[H]
    \centering
    \begin{tabular}{|c|c|}
    \hline
        Neural network method & Number of trainable parameters\\
        \hline
        ResNet\cite{He_2016_CVPR}&$>$0.27M\\
        \hline
        Inception V3/ GoogLeNet\cite{Andersen:19,Szegedy2015}&23.6M\\
        \hline
        Xception\cite{Khorin_2021,Chollet2016}&22.8M\\
        \hline
        Deep Image Prior\cite{Ulyanov_2020}&2M\\
        \hline
        PHASENET\cite{Saha:20,Saha:20:github}&1M\\
        \hline
        MLAO in this paper&0.028M to 0.032M\\
        \hline
    \end{tabular}
    \caption{A list of NNs used in image processing and phase determination with their number of trainable parameters. Inception V3\cite{Andersen:19}, Xception\cite{Khorin_2021} and PHASENET\cite{Saha:20} have been directly demonstrated for phase determination. ResNet is a common basic NN architecture that has been used in many different image processing and phase determination architectures\cite{Saha:20}. A 20 layer ResNet is the smallest architecture proposed in the ResNet paper\cite{He_2016_CVPR} that has $\sim$0.27M trainable parameters. Deep Image Prior employs a U-Net architecture that is a commonly used in many biomedical image processing applications. Deep phase decoder\cite{Bostan:20}, a network designed for wavefront and image reconstruction, was also inspired and adapted from Deep Image Prior.}
    \label{tab:NN_parameter_size_comparison}
\end{table}

\section{Choice of bias mode}

\begin{figure}[H]
    \centering
    \includegraphics[width = 1.0\textwidth]{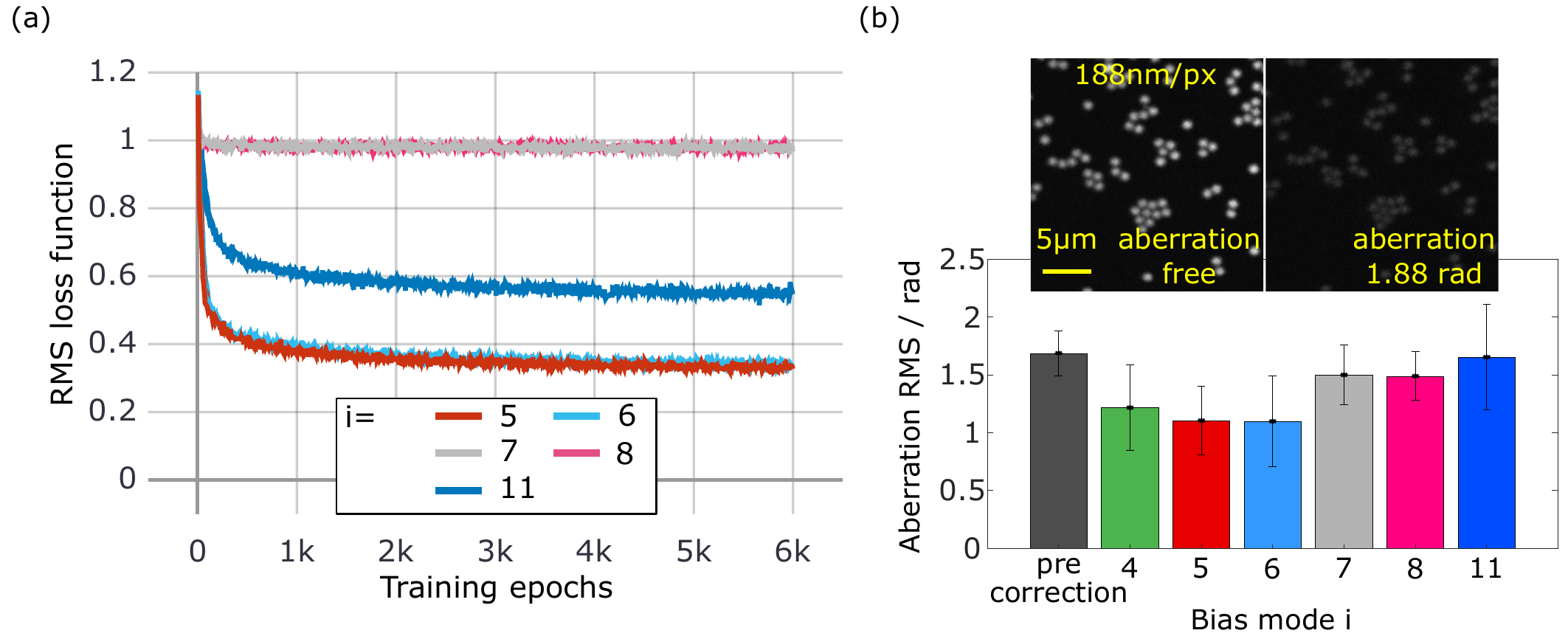}
    \caption{Testing Zernike modes as choice of bias aberration. (a) A plot of the root mean square (RMS) loss function against the number of epochs when training NNs of the same architecture from the same dataset but using different bias modes. (b) Statistical results of testing the trained NNs to correct the same sets of random aberrations over 2-P microscope images of beads. Twenty randomly generated aberrations consisting five Zernike modes and RMS value smaller than 2.2 radians were introduced for correction (dark gray bar). The remaining aberrations after correction by different networks were averaged and shown in the figure; standard deviations of the remaining aberrations are represented as the error bar. Insets showed an example of the FOV when no aberration was introduced and an example when 1.88 rad of aberration was introduced into the system.}
    \label{fig:bias_comparison}
\end{figure}

The simplest MLAO implementation uses a pair of biased images as the input. The nature of the bias aberrations is a design choice. In order to investigate this, we tested individual Zernike modes as the bias and trained different MLAO networks with identical architecture to correct the same randomly generated aberrations. The loss function of the different NNs during training was shown in Fig. \ref{fig:bias_comparison} (a). Results from correcting 20 randomly generated aberrations were shown in Fig. \ref{fig:bias_comparison} (b).

The two networks using oblique and vertical astigmatism (index $i=$5 and 6) converged to similar loss function during training (Fig. \ref{fig:bias_comparison} (a)). The same two networks also gave similar averaged remaining aberrations during experimental aberration correction on a bead sample (Fig. \ref{fig:bias_comparison} (b)). The two networks using vertical and horizontal coma (index 7 and 8) also showed mutually similar values. This was expected as these pairs of modes (5 and 6; 7 and 8) differ only by rotation, which should not have an effect on how effective the networks determine aberrations.

From these results, the NNs using astigmatism as the bias modes converged to the smallest loss function during training. This possibly suggested that the astigmatism modes, on average, allowed the network to learn more from the training data. It was also observed from the experimental results where, in general, the NN obtained the smallest remaining aberrations. We therefore chose to use astigmatism as the modulation modes for the two-bias NN methods in the experiments conducted in this paper.




\section{Tolerance to sampling rate}
\begin{figure}[H]
    \centering
    \includegraphics[width = 0.6\textwidth]{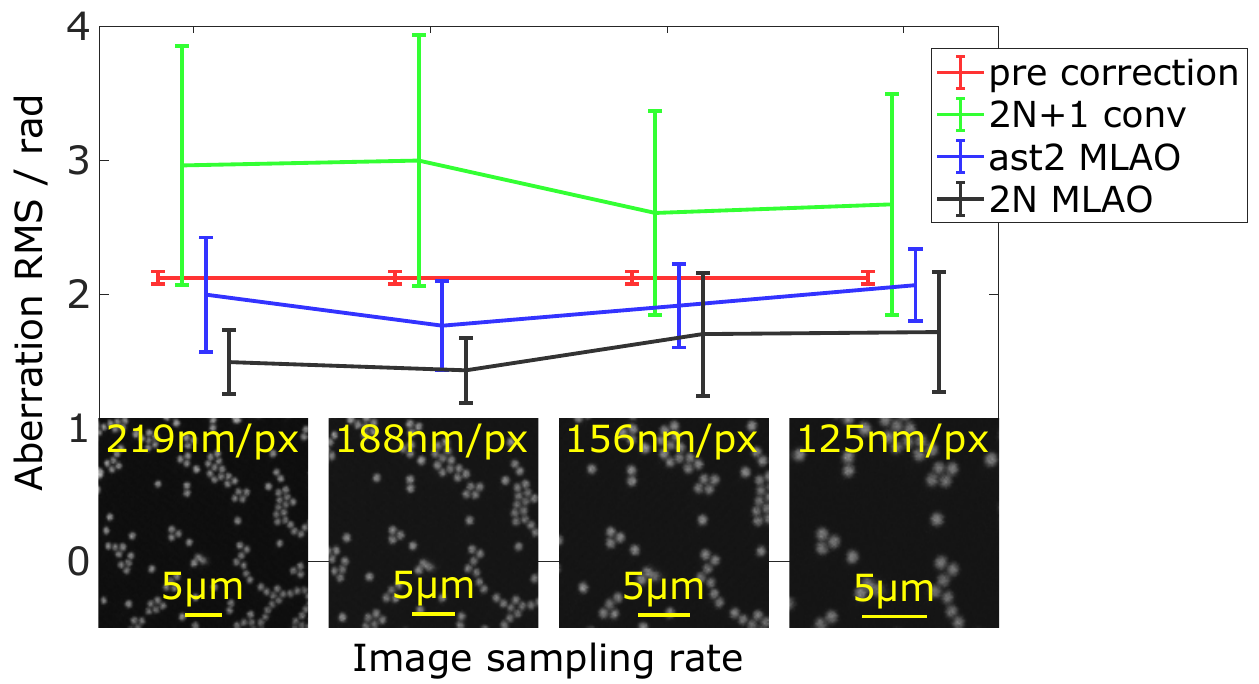}
    \caption{Testing of robustness to pixel sampling. Statistical results of remaining aberrations before (red plot) and after correction using \emph{2N+1 conv}, \emph{ast2 MLAO} and \emph{2N MLAO} methods. The results were averaged from 20 randomly generated aberrations and the SDs were shown as the error bars. The same algorithms were used to correct the same aberrations over images collected at different pixel sampling as shown by the horizontal axis. Insets show examples of the images collected at different sampling rates.}
    \label{fig:results_1_ite_beads_varying_sampling_rate}
\end{figure}

As described in the paper, the networks for scanning microscopy were trained on simulated dataset with pixel sampling within the range of $1.0\times$ to $1.2\times$ of the base sampling rate (see the method section in the main paper for more details). However in many practical cases, there can be uncertainty in pixel sampling for a system or constraints on the sampling rates that may be used. We hence tested the tolerance of our networks to pixel sampling rates outside the range of the training dataset (see Fig. \ref{fig:results_1_ite_beads_varying_sampling_rate}).

In this case, 188nm per pixel was close to the sampling of the generated dataset on which the two NNs were trained. When images were sampled at a smaller or larger rate, \emph{ast2 MLAO} and \emph{2N MLAO} were still able to correct aberrations, but were slightly less effective.

\section{Further microscope demonstrations}
\begin{figure}
    \centering
    \includegraphics[width = 1.0\textwidth]{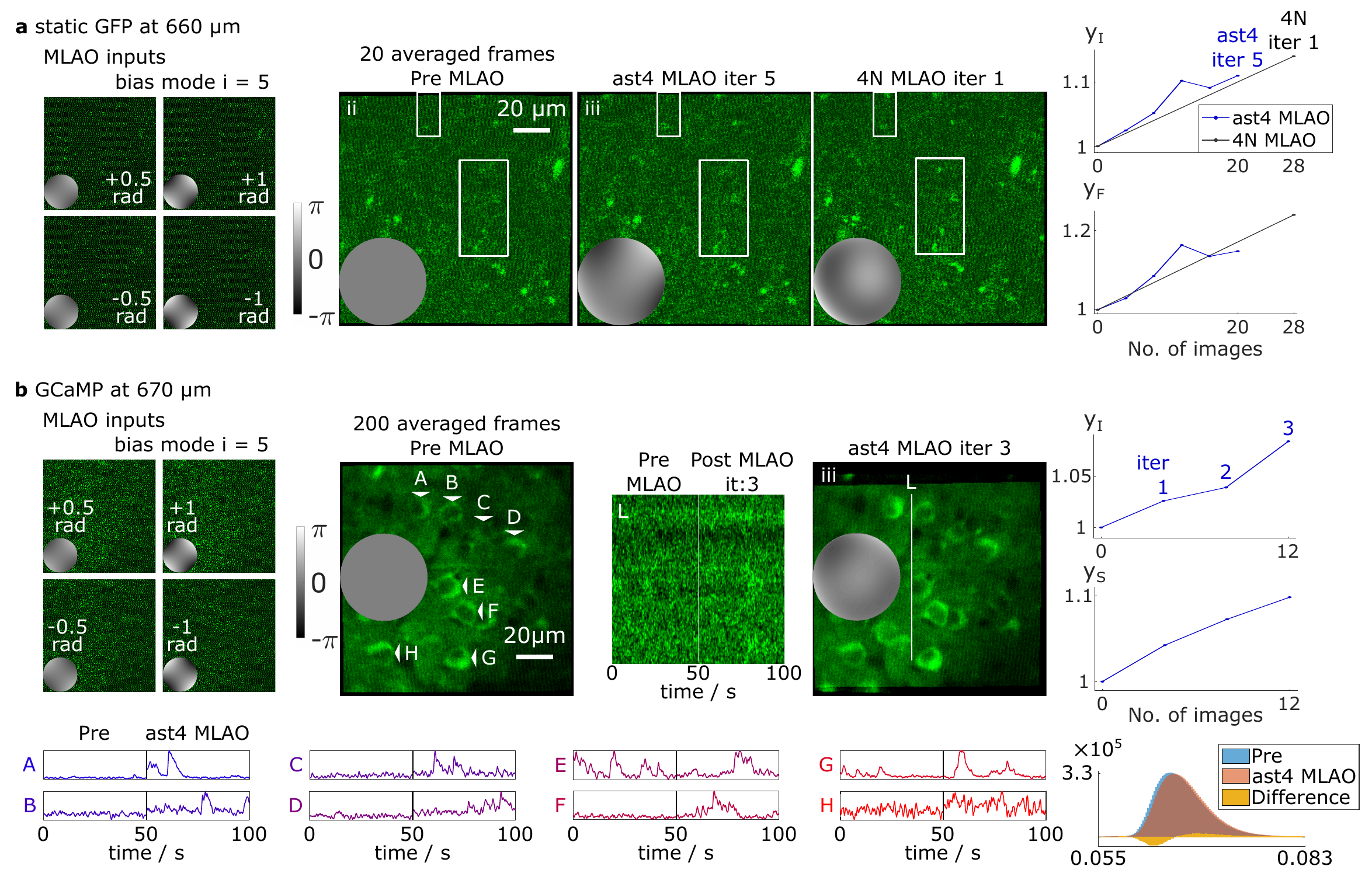}
    \caption{Three-photon microscopy imaging static GFP at $660\mu m$ and GCaMP neuronal activities at depth $670\mu m$. Power at sample was 32.33 mW and 44 mW respectively. Wavefronts inserted to the figures showed the phase modulations applied by the DM at the relevant step; the common scale for each set of results is indicated by the grayscale bars in (a) and (b).
    (a) shows on the left example single-frame images used in correction with the corresponding bias modes as insets; these were the image inputs to \emph{ast4 MLAO}. For \emph{4N MLAO}, six more bias modes and thus 24 more images were also used in each iteration.
    Three images at the central panel are shown averaged from 20 frames after motion correction. The rectangular boxes highlight regions of interest for comparison.
    The plots on the right show the intensity metric ($\text{y}_\text{I}$) and the Fourier metric ($\text{y}_\text{F}$), respectively, calculated from single image frames, against the number of images acquired for five correction iterations of \emph{ast4 MLAO} one correction iteration of \emph{4N MLAO}.\\
    (b) shows on the left example single-frame images used as inputs to the \emph{ast4 MLAO} correction with the corresponding bias modes as insets.
    The central panel shows respectively before and after \emph{ast4 MLAO} correction through three iterations, 200 frame averages after motion correction.
  The time traces were taken from the marked line L.
    The plots on the right show the intensity metric ($\text{y}_\text{I}$) and the sharpness metric ($\text{y}_\text{S}$), respectively, calculated from single image frames, against the number of images acquired for five iterations \emph{ast4 MLAO}.
    The lower panel shows the calcium activity of 8 cells (A-H marked on the averaged image). The lower right plot shows the histograms of the 200 frames collected before (blue) and after (red) \emph{ast4 MLAO} corrections. The difference between the after and before \emph{ast4 MLAO} is marked in yellow. The pixel values are normalised between 0 and 1.
    }
    \label{fig:results-3-p-gcamp-670um}
\end{figure}
 Figure \ref{fig:results-3-p-gcamp-670um} showed the performance of the \emph{4N MLAO} and \emph{ast4 MLAO} algorithm, for imaging GFP labelled processes and neuronal activity at a depth of more than 600 $\mu$m in a mouse brain when using a 3-P system. Despite the very low SNR of the image data, the image quality and cell activity data were considerably improved.

 \begin{figure}
    \centering
    \includegraphics[width = 1.0\textwidth]{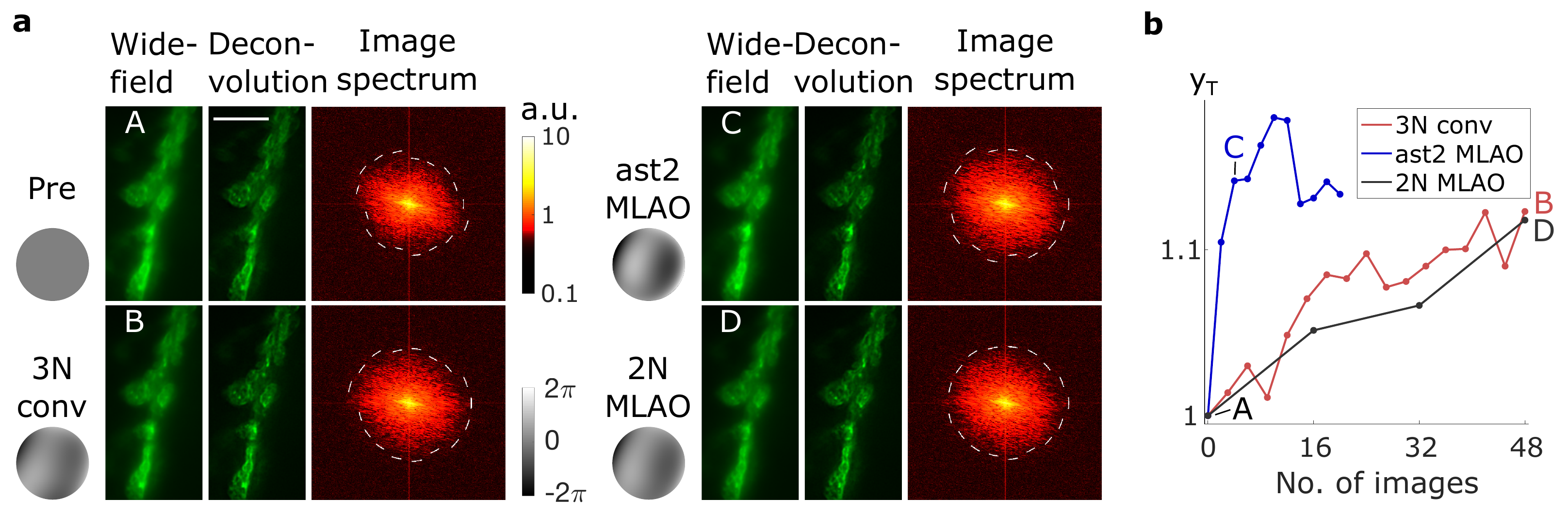}
    \caption{Aberration correction in a widefield 3-D structured illumination microscope (SIM).
    (a) Widefield images acquired A before and B-D after correction through different methods (as marked on the metric plot (b)). The first column shows wavefronts corrected by the DM for each image acquisition; phase is shown on the adjacent scale bar. The second column shows the widefield images and the third column shows corresponding deconvolved widefield images. 
The fourth column shows corresponding image spectra of the second column widefield images displayed in a logarithmic scale (as shown in the colorbar); dashed lines show the threshold where signal falls below the noise level.\\
    (b) The frequency threshold metric $\text{y}_\text{T}$ against the number of images, for two iterations of \emph{3N conv}, ten iterations of \emph{ast2 MLAO} and three iterations of \emph{2N MLAO}.}
    \label{fig:Supplemental_widefield_results}
\end{figure}
 Figure \ref{fig:Supplemental_widefield_results} showed the performance of the \emph{ast2 MLAO} and \emph{2N MLAO} algorithm when imaging neural muscular junctions in a single photon widefield system. The results showed that \emph{ast2 MLAO} corrected much faster than the other two methods.

\section{Details of the experimental methodology}\label{section:method:experimental_setup_sample_preparation_network_parameters}
Three optical systems, a 2-P, 3-P and widefield microscope, were used for demonstrations on different samples. Networks with different parameter settings are also adjusted for different applications.

\subsection{Experimental setups}
\begin{figure}
    \centering
    \includegraphics[width = 0.75\textwidth]{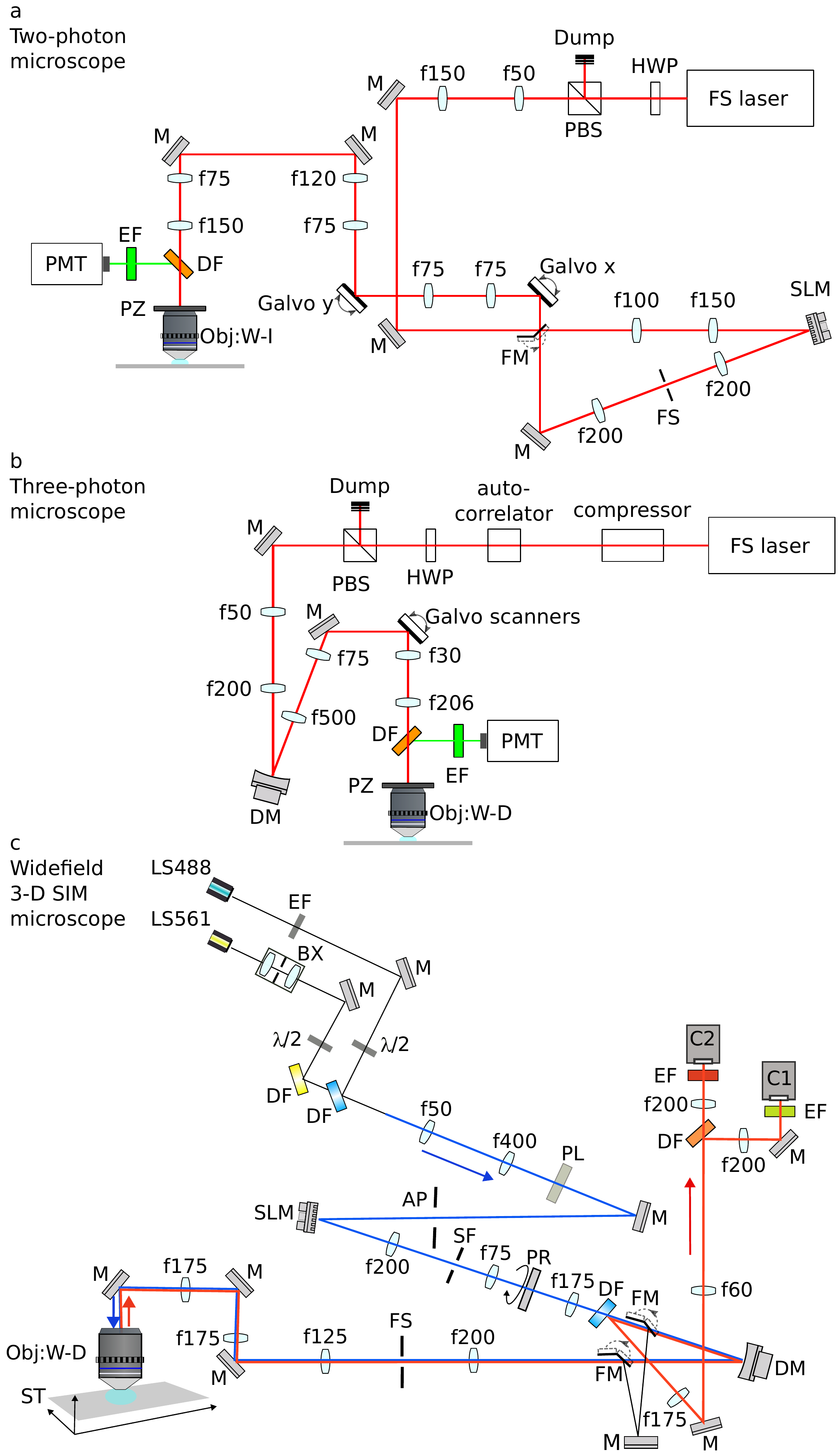}
    \caption{Configuration of the (a) 2-P (b) 3-P (c) widefield 3-D SIM microscope. Femtosecond (FS) Laser; Continuous-wave lasers with wavelenths 488nm and 561nm (LS488 and LS561);
 half wave plate (HWP); polarisation beam splitter (PBS); laser beam dump (Dump); lens with focal length $= x$ mm (f$x$); broadband dielectric mirror (M); flip mirror (FM); Hamamatsu spatial light modulator (SLM); Mirao 52E deformable mirror (DM) in the 3-P system; ALPAO 69 deformable mirror (DM) in the widefield 3-D SIM system; aperture (AP); spatial filter (SF); field stopper (FS); X galvanometer (Galvo x); Y galvanometer (Galvo y); beam expansion (BX); half waveplate ($\lambda/2$); linear polariser (PL); polarisation rotator (PR); Olympus $40\times$ numerical aperture (NA) 1.15 water immersion objective lens (Obj:W-I) used in the 2-P system; Nikon $16\times$ NA 0.8 water dipping objective lens (Obj:W-D) used in the 3-P system; Olympus $60\times$ NA 1.1 water dipping objective lens (Obj:W-D) in the widefield 3-D SIM system; Z-piezo translation stage (PZ); X-Y-Z translational sample mounting stage (ST); Dichroic filter (DF) allow emission signal from fluorophores to be reflected through emission filter (EF) into a photo-multiplier tube (PMT) in a multi-photon system; cameras (C1 and C2)}
    \label{fig:2-P-3-P-widefield-setup}
\end{figure}
\subsection{Sample preparation}

The 3-P results were collected from imaging male (Lhx6-eGFP)BP221Gsat; Gt(ROSA)26Sortm32(CAG-COP4*H134R/EYFP)Hze mice (static imaging) and female and male Tg(tetO-GCaMP6s)2Niell mice (calcium imaging). Mice were between 8-12 weeks of age when surgery was performed. The scalp was removed bilaterally from the midline to the temporalis muscles, and a metal headplate with a 5 mm circular imaging well was fixed to the skull with dental cement (Super-Bond C\&B, Sun-Medical). A 4–5 mm circular craniotomy was performed during which any bleeding was washed away with sterile external solution or staunched with Sugi-sponges (Sugi, Kettenbach). Cranial windows composed of 4 or 5 mm circular glass coverslips were press-fit into the craniotomy, sealed to the skull by a thin layer of cyanoacrylate (VetBond) and fixed in place by dental cement.

The widefield 3-D SIM results were collected from imaging NMJ of \textit{Drosophila} larvae. For the immunofluorescence sample with one coloured channel, it was prepared as previously \cite{Brent:2009aa}. Crawling 3rd instar larvae of wildtype Oregon-R \textit{Drosophila melanogaster} were dissected on a Sylgard-coated Petri Dish in HL3 buffer with 0.3mM Ca2+ to prepare larval fillet \cite{Parton2010}. Then, the larval fillet samples were fixed in Paraformaldehyde 4\% in PBS containing 0.3\% (v/v) Triton X-100 (PBSTX) for 30 minutes. The brains were removed post-fixation, and the fillet samples were transferred to a Microcentrifuge tube containing PBSTX for 45 minutes of permeabilisation. The samples were stained with HRP conjugated to Alexa Fluor 488 and DAPI for 1 hour at room temperature ($21C^{\circ}$). After the washes, the samples were mounted in Vectashield.

For the 3-D SIM results collected on the \textit{Drosophila} larvae sample with two coloured channels, it was prepared by following the protocol presented in \cite{Brent:2009aa}. 3rd instar \textit{Drosophila} melanogaster larvae (Brp-GFP strain) were dissected in HL3 buffer with 0.3mM Ca2+ to prepare a so-called larval fillet, and the larval brains were removed. After this, larvae were stained for 15 minutes with HRP conjugated to Alexa Fluor 568 to visualise the neurons, washed with HL3 buffer with 0.3mM Ca2+ and imaged in HL3 buffer without Ca2+ to prevent the larvae from moving.

\subsection{Network parameters}\label{section:Supplimental:network_parameters}
Table \ref{tab:MLAO_parameters} showed the network settings used in different imaging applications.

\begin{table}[H]
    \centering
    \begin{tabular}{|c|c|c|c|c|c|c|}
    \hline
        Results in&Method label&M&N&Bias&Bias&Corrected\\
        &&&&modes, i&depths&modes, i\\
        \hline
        Fig. 2 (a, c, f)&\emph{ast2 MLAO} &2&5& 5&$\pm 1$ rad& 5\textendash8, 11\\
        Fig. \ref{fig:results_1_ite_beads_varying_sampling_rate}&&&&&&\\
        \hline
        Fig. 2 (a, c, f)&\emph{2N MLAO}&10&5& 5\textendash8, 11&$\pm 1$ rad& 5\textendash8, 11\\
        Fig. \ref{fig:results_1_ite_beads_varying_sampling_rate}&&&&&&\\
        \hline
        Fig. 2 (b, d, e)&\emph{ast2 MLAO}&2&9& 5&$\pm 1$ rad& 5\textendash13\\
        \hline
        Fig. 2 (b, d, e)&\emph{2N MLAO}&18&9& 5\textendash13&$\pm 1$ rad& 5\textendash13\\
        \hline
        Fig. 3 (a, b)&\emph{ast4 MLAO}&4&7& 5&$\pm 0.5$& 5\textendash11\\
        Fig. \ref{fig:results-3-p-gcamp-670um}&&&&&$\pm 1$ rad&\\
        \hline
        Fig. 3 (a)&\emph{4N MLAO}&28&7& 5\textendash11&$\pm 0.5$& 5\textendash11\\
        Fig. \ref{fig:results-3-p-gcamp-670um}&&&&&$\pm 1$ rad&\\
        \hline
        Fig. 4 &\emph{ast2 MLAO}&2&8& 5&$\pm 1$ rad& 5\textendash11, 22\\
        Fig. \ref{fig:Supplemental_widefield_results}&&&&&&\\
        \hline
        Fig. 4 &\emph{2N MLAO}&2&8& 5\textendash11, 22&$\pm 1$ rad& 5\textendash11, 22\\
        Fig. \ref{fig:Supplemental_widefield_results}&&&&&&\\
        \hline
    \end{tabular}
    \caption{A list of MLAO parameters chosen for different imaging applications. The Zernike modes were sequenced using Noll's indices.}
    \label{tab:MLAO_parameters}
\end{table}

\bibliography{sample_sup}
